\documentclass[preprintnumbers,nofootinbib,10pt,superscriptaddress]{revtex4-2}
\usepackage[T1]{fontenc}
\usepackage[latin9]{inputenc}
\usepackage{geometry}
\usepackage[active]{srcltx}
\usepackage{color}
\usepackage{babel}
\usepackage{mathrsfs}
\usepackage{amsmath,bm}
\usepackage{amssymb}
\usepackage{graphicx}
\usepackage{ulem}
\usepackage[unicode=true,pdfusetitle,
 bookmarks=true,bookmarksnumbered=false,bookmarksopen=false,
 breaklinks=false,pdfborder={0 0 0},pdfborderstyle={},backref=false,colorlinks=true]
 {hyperref}
\hypersetup{
 linkcolor=blue, citecolor=cyan}
\makeatletter
\usepackage{slashed}

\allowdisplaybreaks[3]

\usepackage{xcolor}

\makeatother

\begin{document}
\title{Quantum kinetic theory for dynamical spin polarization from QED-type interaction}
\author{Shuo Fang}
\email{fangshuo@mail.ustc.edu.cn}

\affiliation{Department of Modern Physics, University of Science and Technology
of China, Anhui 230026, China}
\author{Shi Pu}
\email{shipu@ustc.edu.cn}

\affiliation{Department of Modern Physics, University of Science and Technology
of China, Anhui 230026, China}
\author{Di-Lun Yang}
\email{dilunyang@gmail.com}

\affiliation{Institute of Physics, Academic Sinica, Taipei 11529, Taiwan}
\begin{abstract}
We investigate the dynamical spin polarization of a massless electron probing an electron plasma in locally thermal equilibrium via the Moller scattering from the quantum kinetic theory. We derive an axial kinetic equation delineating the dynamical spin evolution in the presence of the collision term with quantum corrections up to $\mathcal{O}(\hbar)$ and the leading-logarithmic order in coupling by using the hard-thermal-loop (HTL) approximation, from which we extract the spin-polarization rate induced by the spacetime gradients of the medium.
When the electron probe approaches local equilibrium, we further simplify the collision term into a relaxation-time expression. 
Our kinetic equation may be implemented in the future numerical simulations for dynamical spin polarization. 
\end{abstract}
\maketitle

\section{Introduction\label{sec:Introduction}}
In off-central heavy ion collisions, a large global angular momentum is produced, which may result in the spin polarization of the quark gluon plasma (QGP) from spin-orbit coupling and yield the spin polarization or spin alignment of emitted hadrons \cite{Liang:2004ph,Liang:2004xn,Gao:2007bc,Becattini:2013vja}. Such phenomena have been observed in recent STAR and ALICE experiments \cite{STAR:2017ckg,STAR:2019erd,ALICE:2019aid,Singha:2020qns}.
Based on the modified Cooper-Frye formula for spin polarization dictated by thermal vorticity in global equilibrium \cite{Becattini:2013fla,Fang:2016vpj}, the global spin polarization of $\Lambda$ (and $\overline{\Lambda}$) hyperons
at intermediate energies can be described by many transport
models \cite{Li:2017slc,Wei:2018zfb,Karpenko:2016jyx,Xie:2017upb,Shi:2017wpk,Fu:2020oxj}. See also Ref.~\cite{Sun:2017xhx} from the non-equilibrium kinetic-theory approach and Ref.~\cite{Ryu:2021lnx} with the inclusion of corrections in local equilibrium. There have been further measurements for the spin polarization of $\Lambda$ hyperons in collisions at low energies \cite{STAR:2021beb,Kornas:2020qzi} and related studies in theory \cite{Guo:2021udq,Ivanov:2020udj,Deng:2020ygd,Deng:2021miw}. Moreover, there are also theoretical studies attempting to explain the spin alignment of vector mesons \cite{Sheng:2019kmk,Sheng:2020ghv,Xia:2020tyd,Muller:2021hpe,Yang:2021fea} and the proposal for modifications on the yields of hadrons with different spin due to vorticity \cite{ExHIC-P:2020tcv}.

However, the theoretical description with the global equilibrium assumption of local spin polarization in
the longitudinal direction does not match the experimental result measured in Au-Au collisions at $\sqrt{s_{NN}}=200$ GeV by STAR \cite{STAR:2019erd}. The models based
on hydrodynamics and transport theories predict an opposite trend
of the longitudinal spin polarization as a function of azimuthal angle
\cite{Becattini:2017gcx,Xia:2018tes}. The contradiction is also known as the ``sign
problem'' for local spin polarization. It is realized that the spin polarization in global equilibrium may not resolve the problem and further corrections beyond global equilibrium should be considered. There have been several studies to address this issue \cite{Voloshin:2017kqp,Liu:2019krs,Becattini:2019ntv,Xia:2019fjf,Li:2021zwq,Wu:2019eyi,Wu:2020yiz,Fu:2021pok,Becattini:2021iol,Yi:2021ryh,Florkowski:2021xvy,Sun:2021nsg}. In particular, the so-called thermal shear correction in local equilibrium, obtained from the linear response theory \cite{Liu:2020dxg,Liu:2021uhn}, statistical field theory \cite{Becattini:2021suc}, and chiral kinetic theory (CKT) for massless fermions \cite{Hidaka:2017auj}, could yield substantial contribution to the longitudinal spin polarization. The inclusion of this shear correction may successfully describe the experimental measurement with certain approximations as shown by hydrodynamic simulations \cite{Fu:2021pok,Becattini:2021iol}. Nonetheless, the numerical results could be sensitive to the chosen parameters and adopted approximations \cite{Yi:2021ryh,Florkowski:2021xvy,Sun:2021nsg, Wu:2022mkr}. The shear corrections have also been studied in the helicity polarization \cite{Yi:2021unq}, which could provide a baseline for the local polarization led by vorticity and a probe for the initial axial chemical potential \cite{Becattini:2020xbh,Gao:2021rom}. Furthermore, when considering the local-equilibrium condition, the dissipative corrections pertinent to interaction should also be involved.  
More discussions and details can be found in recent reviews \cite{Wang:2017jpl,Becattini:2020ngo,Becattini:2020sww,Gao:2020vbh,Liu:2020ymh,Becattini:2022zvf}.


Nowadays there are two primary approaches to explore dynamical spin polarization and the dissipative effects. One is the spin hydrodynamics
based on the conservation laws 
\cite{Hattori:2019lfp,Fukushima:2020qta,Fukushima:2020ucl,Li:2020eon,She:2021lhe,Montenegro:2017lvf,Montenegro:2017rbu,Florkowski:2017ruc,Florkowski:2018myy,Bhadury:2020puc,Shi:2020qrx,Becattini:2018duy,Gallegos:2021bzp,Hongo:2021ona,Florkowski:2017dyn,Florkowski:2018ahw,Florkowski:2019qdp,Florkowski:2019voj,Bhadury:2020cop,Shi:2020htn,Singh:2020rht,Florkowski:2021wvk,Wang:2021ngp,Wang:2021wqq,Liu:2020ymh,Hongo:2022izs} (see also Ref.~\cite{Florkowski:2018fap} for a review). 
One can construct spin hydrodynamics using the entropy principle  \cite{Hattori:2019lfp,Fukushima:2020qta,Fukushima:2020ucl,Li:2020eon,She:2021lhe},
the effective Lagrangian theory \cite{Montenegro:2017lvf,Montenegro:2017rbu},
the kinetic-theory approach \cite{Yang:2018lew,Florkowski:2017ruc,Florkowski:2018myy,Florkowski:2018fap,Bhadury:2020puc,Shi:2020qrx,Weickgenannt:2022zxs},
and quantum field theory \cite{Becattini:2018duy,Gallegos:2021bzp,Hongo:2021ona,Hongo:2022izs}. In general, the spin hydrodynamics is a macroscopic effective theory including spin effects
and the conservation of angular momentum as an extension of standard relativistic hydrodynamics.
The other is quantum
kinetic theory (QKT) as a microscopic spin transport theory in connection to underlying quantum field theories \cite{Gao:2019znl,Weickgenannt:2019dks,Weickgenannt:2020aaf,Hattori:2019ahi,Yang:2020hri,Liu:2020flb,Weickgenannt:2021cuo,Sheng:2021kfc,Wang:2021qnt,Huang:2020wrr,Wang:2020dws,Weickgenannt:2022jes} (see Refs.~\cite{Gao:2020vbh, Gao:2020pfu, Hidaka:2022dmn} for recent reviews).
The QKT is an extension of the CKT for massless fermions \cite{Gao:2012ix,Son:2012wh,Son:2012zy,Stephanov:2012ki,Chen:2013iga,Chen:2014cla,Chen:2015gta,Chen:2012ca,Hidaka:2016yjf,Huang:2018wdl,Mueller:2017arw,Mueller:2017lzw,Manuel:2013zaa,Manuel:2014dza,Carignano:2018gqt,Carignano:2019zsh,Lin:2018aon,Lin:2019ytz,Carignano:2021zhu}.
It can be consistently derived from the covariant Keldysh formalism and Wigner-function
method from quantum field theory.

To understand the dynamical spin polarization in QKT, it is inevitable to incorporate the quantum corrections on collisions. In recent years, there have been intensive studies along this direction \cite{Li:2019qkf,Yang:2020hri,Weickgenannt:2020aaf,Hattori:2020gqh,Weickgenannt:2021cuo,Sheng:2021kfc,Lin:2021mvw,Wang:2020pej,Wang:2021qnt,Hongo:2022izs,Sheng:2022ssd,Das:2022azr}. Nevertheless, most of studies consider effective models rather than the gauge theory for simplicity. In Refs.~\cite{Li:2019qkf,Yang:2020hri,Hongo:2022izs}, only part of the collision term giving rise to spin relaxation is computed in weakly coupled quantum chromodynamics (QCD). As a toy model for studying dynamical spin polarization in the QGP\footnote{In principle, the ultimate goal is to study the strange quark probing QGP composed of massless quarks and gluons in equilibrium. Nevertheless, there have not been sufficient understanding for the quantum corrections of polarized gluons even in thermal equilibrium (see Refs.~\cite{Huang:2020kik,Hattori:2020gqh,Lin:2021mvw,Mameda:2022ojk} for some recent studies for polarized photons and QKT). Technically, it is also more involved to work with QKT for massive fermions.}, our work here is to investigate the quantum-electrodynamics (QED)-type interaction by considering 2-2 scattering process for massless fermions in the absence of onshell photons. 
In our theoretical setup, we emit a probe electron to interact with an electron plasma in thermal equilibrium and study the
spin polarization of the probe. We calculate collision kernels of QKT up to the leading logarithmic order in electric coupling $e$ and to $\mathcal{O}(\hbar)$ as the leading-order quantum correction using the HTL approximation following the procedure in Ref.~\cite{Yang:2020hri}. Such quantum corrections result in the spin-polarization rate pertinent to the gradient terms of the medium in local equilibrium, which includes thermal vorticity, shear strength, and the gradient of the ratio of a chemical potential to temperature. When the probe electron is close to local equilibrium, we further derive a simplified collision term using the relaxation-time approximation (RTA) with the (inverse) relaxation times
in operator form \footnote{There has been a similar approach for studying chiral effects of neutrino transport by CKT in core-collapse supernovae \cite{Yamamoto:2020zrs,Yamamoto:2021hjs}.}. 
A similar study of massive fermions in the Nambu-Jona-Lasinio (NJL) model has been reported in Ref.~\cite{Wang:2021qnt}, whereas only the result with a medium in global equilibrium is considered.

This paper is organized as follows: In Sec.\ref{sec:Chiral-kinetic-equation},
we briefly review the Wigner-function approach and derivation of the master
equations giving rise to the QKT based on the Keldysh formalism for the massless fermions with the power-counting
scheme in \cite{Yang:2020hri}. In Sec.\ref{sec:general_setup}, we introduce the general setup for our QKT of an electron probe interacting with the medium and expatiate how we handle the collision term. In Sec.\ref{sec:Collision-kernels-in},
we compute the collision kernel up to the leading-logarithmic order by using the HTL approximation and assuming the medium is in local equilibrium. From the collision term with quantum corrections, the spin-polarization rate is found. A brief summary and discussions are presented. In Sec.\ref{sec:relaxation_times},
we further derive the simplified kinetic equation assuming the electron probe is near local equilibrium by using the RTA and extract the interaction-dependent relaxation
times in operator form. We finally conclude our results and make an 
outlook in Sec.\ref{sec:Conclusions-and-outlook}. Some critical steps in derivations are presented in Appendices.

We adopt the Minkowski spacetime metric, $g^{\mu\nu}=g_{\mu\nu}=\mathrm{diag}(+1,-1,-1,-1)$, and the Dirac matrices $\gamma^{\mu}$
in the Weyl basis. 
We introduce $\sigma_{\mu}=(1,\bm{\sigma})$ and $\overline{\sigma}_{\mu}=(1,-\bm{\sigma})$
with $\sigma_i$ the Pauli matrices and $\gamma_{5}=i\gamma^{0}\gamma^{1}\gamma^{2}\gamma^{3}$. The Levi-Civita
symbol is chosen as 
$\epsilon^{0123}=-\epsilon_{0123}=+1.$
We denote $A_{(\mu\nu)}=A_{\mu\nu}+A_{\nu\mu}$ and $A_{[\mu\nu]}=A_{\mu\nu}-A_{\nu\mu}$. 
We have also used the notation for the projector
$\Delta^{\mu\nu}=g^{\mu\nu} - u^{\mu}u^{\nu}$ and $\Theta^{\mu\nu}(p)=g^{\mu\nu}-u^{\mu}u^{\nu}+\hat{p}^{\mu}_{\perp}\hat{p}^{\nu}_{\perp}$ with $u^\mu$ being the fluid velocity, $p^{\mu}_{\perp}\equiv\Delta^{\mu\nu}p_\nu$ and $\hat{p}^{\mu}_{\perp}\equiv p^\mu_\perp/\sqrt{-p_\perp \cdot p_\perp}$.

\section{Quantum kinetic theory for massless fermions \label{sec:Chiral-kinetic-equation}}

We start from the standard QED Lagrangian for massless fermions in the Weyl basis,
\begin{eqnarray}
\mathcal{L} 
  =  \psi_{L}^{\dagger}\overline{\sigma}^{\mu}i\hbar D_{\mu}\psi_{L}+\psi_{R}^{\dagger}\sigma^{\mu}i\hbar D_{\mu}\psi_{R},\label{eq:QED_Massless_L_2}
\end{eqnarray}
where $D_{\mu}=\partial_{\mu}+iqeA_{\mu}/\hbar$ and the left-
and right-handed fermions are disentangled. We will set $q=-1$ for electrons in the following context. Taking the ensemble average, we define the lessor and greater two-point Green function for Weyl fermions, 
\begin{equation}
S_{L/R}^{<}(x,y)=\langle U(y,x)\psi_{L/R}^{\dagger}(y)\psi_{L/R}(x)\rangle,\;\;S_{L/R}^{>}(x,y)=\langle U(y,x)\psi_{L/R}(x)\psi_{L/R}^{\dagger}(y)\rangle,\label{eq:Lessor/Greater_GF_Definition}
\end{equation}
where
$U(x,y)=\exp\left(-\frac{iqe}{\hbar}\int_{y}^{x}dz\cdot A(z)\right)$ represents the gauge link to maintain the gauge invariance. We then introduce the Wigner transformation 
\begin{equation}
S^{\lessgtr}_{L/R}(p,X)=\int\frac{d^{4}Y}{(2\pi\hbar)^{4}}e^{\frac{i}{\hbar}p\cdot Y}S_{L/R}^{\lessgtr}(X+\frac{Y}{2},X-\frac{Y}{2}),\label{eq:Wigner Transformation}
\end{equation}
where $X=\frac{x+y}{2}$ and $Y=x-y$. We will then focus on the right-handed fermions and the formalism for left-handed fermions can be analogously derived. Based on the Dyson-Schwinger equation and Dirac equation, one can derive the Kadanoff-Baym equations up to $\mathcal{O}(\hbar)$, which includes the leading-order quantum correction \cite{Blaizot:2001nr,Hidaka:2016yjf,Hidaka:2022dmn},
\begin{eqnarray}\label{eq:kinetic_collision}
	\sigma^{\mu}\left(p_{\mu}+\frac{1}{2}i\hbar\Delta_{\mu}\right)S^{<}_R & = & \frac{i\hbar}{2}(\Sigma^{<}_RS^{>}_R-\Sigma^{>}_RS^{<}_R),\\
	\left(p_{\mu}-\frac{1}{2}i\hbar\Delta_{\mu}\right)S^{<}_R\sigma^{\mu}& = & -\frac{i\hbar}{2}(S^{>}_R\Sigma^{<}_R-S^{<}_R\Sigma^{>}_R),
\end{eqnarray}
where $\Delta_{\mu}=\partial_{X\mu}+eF_{\mu\nu}\partial_{p}^{\nu}$, and $\Sigma^{<}_R$ and $\Sigma^{>}_R$ denote the lesser and greater self-energies for right-handed fermions. \footnote{Roughly speaking, $\Sigma^<_{R}$ and $\Sigma^>_{R}$ are proportional to the emission and absorption rates of the medium that contribute to the gain and loss of the probe, respectively. In the 2 to 2 scattering as will be elaborated later, $\Sigma^<_R$ is proportional to the distribution functions of one outgoing particle and two incoming particles and vice versa for $\Sigma^>_R$.} Here we have dropped the one-particle potential and real parts of retarded self-energy and Wigner function since they do not directly affect the collisional effect of our interest \cite{Hidaka:2016yjf}. Also, we will hereafter neglect background electromagnetic fields by taking $F_{\mu\nu}=0$. 

It is convenient for the follow-up computations by parametrizing $S^{\lessgtr}_R=\overline{\sigma}^{\mu}S^{\lessgtr}_{\mu}$ and $\Sigma^{\lessgtr}_R=\sigma^{\mu}\Sigma^{\lessgtr}_{L\mu}$.
Note that the decomposition of self-energies is different from the decomposition of
the Wigner function and hence we denote $\Sigma_{L}^{\lessgtr\mu}$ here. 
Perturbatively solving Eq.~(\ref{eq:kinetic_collision}) up to $\mathcal{O}(\hbar)$, one obtains the solution of Wigner functions,
\begin{eqnarray}
S_{R}^{\lessgtr,\mu}(p,x)  =  2\pi\mathrm{sgn}(n\cdot p)\delta(p^{2})\Big[p^{\mu}f_{R}^{\lessgtr}(p,x)+\hbar S^{\mu\nu}_{(n)}\left(\partial_{\nu}f_{R}^{\lessgtr}(p,x)-C_{R,\nu}[f_{R}^{\lessgtr}]\right)\Big],
\label{eq:S<_R}
\end{eqnarray}
where 
$S^{\mu\nu}_{(n)}=\epsilon^{\mu\nu\rho\sigma}p_{\rho}n_{\sigma}/(2n\cdot p)$ corresponds to the spin tensor depending on a time-like frame vector $n^{\mu}$ with $n^2=1$ and $C_{R,\nu}[f_{R}^{\lessgtr}]=\Sigma_{L,\nu}^{\lessgtr}f_{R}^{\gtrless}-\Sigma_{L,\nu}^{\gtrless}f_{R}^{\lessgtr}$. Here $n^{\mu}$ originates from the choice of a spin basis and does not affect the physical quantities. Also, $f_{R}^{\lessgtr}(p,x)$ denote the lesser/greater distribution functions for right-handed fermions, which follow the relation $f_{R}^{<}(p,x)+f_{R}^{>}(p,x)=1$. The sign of energy $\mathrm{sgn}(n\cdot p)$ is involved to incorporate both the particle and anti-particle, while we will omit the part for anti-particles in the later computations. The dynamics of $f_{R}^{<}(p,x)$ is dictated by the kinetic equation,
\begin{equation}
\partial\cdot S_{R}^{<}=\Sigma_{L}^{<}\cdot S_{R}^{>}-\Sigma_{L}^{>}\cdot S_{R}^{<}.
\label{eq:CKT_SR}
\end{equation}
Similarly, we can derive the Winger function and kinetic equation for left-handed
fermions,
\begin{eqnarray}
S_{L}^{\lessgtr,\mu}(p,x)  =  2\pi\mathrm{sgn}(n\cdot p)\delta(p^{2})\Big[p^{\mu}f_{L}^{\lessgtr}(p,x)-\hbar S^{\mu\nu}_{(n)}\left(\partial_{\nu}f_{L}^{\lessgtr}(p,x)-C_{L,\nu}[f_{L}^{\lessgtr}]\right)\Big]
\label{eq:S<_L}
\end{eqnarray}
and
\begin{equation}
\partial\cdot S_{L}^{<}=\Sigma_{R}^{<}\cdot S_{L}^{>}-\Sigma_{R}^{>}\cdot S_{L}^{<}.\label{eq:SKT_SL}
\end{equation}

In a physical system when both right- and left-handed fermions co-exist, it would be more convenient to rewrite Eqs.~(\ref{eq:S<_R}-\ref{eq:SKT_SL})
in terms of the axial-vector basis. That is, we may construct the Wigner functions in Eq.~(\ref{eq:Lessor/Greater_GF_Definition}) with massless Dirac fermions $\psi=(\psi_L,\psi_R)^{\rm T}$ such that\footnote{Other components in the basis of Clifford algebra vanish in the massless case.} 
\begin{equation}
	S^{\lessgtr}=\mathcal{V}^{\lessgtr,\mu}\gamma_{\mu}+\mathcal{A}^{\lessgtr,\mu}\gamma^{5}\gamma_{\mu}.
\end{equation}
The vector and axial-vector components are now related to $S_{L/R}^{\lessgtr}$
by the relation
\begin{equation}
S_{R,\mu}^{<}=\mathcal{V}_{\mu}^{<}+\mathcal{A}_{\mu}^{<},\;\;S_{L,\mu}^{<}=\mathcal{V}_{\mu}^{<}-\mathcal{A}_{\mu}^{<}. \label{eq:def_V_A}
\end{equation}
A similar expression is also applied to self-energies,
\begin{equation}
	\Sigma^{\lessgtr}=\Sigma_V^{\lessgtr,\mu}\gamma_{\mu}+\Sigma_A^{\lessgtr,\mu}\gamma^{5}\gamma_{\mu},
\end{equation}
where
\begin{equation}
\Sigma_{R,\mu}^{<}=\Sigma_{V,\mu}^{<}+\Sigma_{A,\mu}^{<},\;\;\Sigma_{L,\mu}^{<}=\Sigma_{V,\mu}^{<}-\Sigma_{A,\mu}^{<}.
\end{equation}
Accordingly, Eqs.~(\ref{eq:S<_R}) and (\ref{eq:S<_L}) can be rewritten as
\begin{eqnarray}
\mathcal{V}^{\lessgtr,\mu}(p) & = & 2\pi \mathrm{sgn}(n\cdot p)\delta(p^{2})\Big[f_{V}^{\lessgtr}(p)+\hbar S^{(n),\mu\nu}\big(\partial_{\nu}f_{A}^{\lessgtr}
-\Sigma_{V,\nu}^{\lessgtr}f_{A}^{\gtrless}+\Sigma_{V,\nu}^{\gtrless}f_{A}^{\lessgtr}\nonumber \\
 &  & +\Sigma_{A,\nu}^{\lessgtr}f_{V}^{\gtrless}-\Sigma_{A,\nu}^{\gtrless}f_{V}^{\lessgtr}\big)\Big]
 \label{eq:Vector_WF}
\end{eqnarray}
and
\begin{eqnarray}
\mathcal{A}^{\lessgtr,\mu}(p) & = & 2\pi \mathrm{sgn}(n\cdot p)\delta(p^{2})\Big[p^{\mu}f_{A}^{\lessgtr}(p)+\hbar S^{(n),\mu\nu}
\big(\partial_{\nu}f_{V}^{\lessgtr}-\Sigma_{V,\nu}^{\lessgtr}f_{V}^{\gtrless}+\Sigma_{V,\nu}^{\gtrless}(p)f_{V}^{\lessgtr}
\nonumber \\
&  & +\Sigma_{A,\nu}^{\lessgtr}f_{A}^{\gtrless}-\Sigma_{A,\nu}^{\gtrless}f_{A}^{\lessgtr}\big)\Big],\label{eq:Axial_WF}
\end{eqnarray}
where we introduce
the vector-charge distribution function $f^{\lessgtr}_{V}=(f^{\lessgtr}_{R}+f^{\lessgtr}_{L})/2$
and the axial-charge distribution function $f^{\lessgtr}_{A}=(f^{\lessgtr}_{R}-f^{\lessgtr}_{L})/2$, which now follow the relations, $f_{A}^{<}+f_{A}^{>}=0$ and $f_{V}^{<}+f_{V}^{>}=1$. For convenience, we may sometimes denote $\chi^<_{V/A}=\chi_{V/A}$, where $\chi$ can be distribution functions, Wigner functions, or self energies if not specified.
On the other hand, Eqs.~(\ref{eq:CKT_SR}) and (\ref{eq:SKT_SL}) become 
\begin{eqnarray}
\partial\cdot\mathcal{V}^{<} & = & \mathcal{C}_V \equiv \Sigma_{V}^{<}\cdot\mathcal{V}^{>}-\Sigma_{V}^{>}\cdot\mathcal{V}^{<}-\Sigma_{A}^{<}\cdot\mathcal{A}^{>}+\Sigma_{A}^{>}\cdot\mathcal{A}^{<},\label{eq:Vector_KE}
\end{eqnarray}
and\begin{eqnarray}
\partial\cdot\mathcal{A}^{<} & = & \mathcal{C}_A \equiv \Sigma_{V}^{<}\cdot\mathcal{A}^{>}-\Sigma_{V}^{>}\cdot\mathcal{A}^{<}-\Sigma_{A}^{<}\cdot\mathcal{V}^{>}+\Sigma_{A}^{>}\cdot\mathcal{V}^{<}.\label{eq:Axial_KE}
\end{eqnarray}

Following the power-counting scheme in Ref.~\cite{Yang:2020hri}, we may assume the chirality imbalance comes from quantum corrections and hence approximate $\mathcal{V}^{\mu}\sim\mathcal{O}(\hbar^{0})$
and $\mathcal{A}^{\mu}\sim\mathcal{O}(\hbar^{1})$, which imply $f_{V}\sim\mathcal{O}(\hbar^{0})$ and $f_{A}\sim\mathcal{O}(\hbar^{1})$. Similar power
counting should also be applied to the self-energies,  $\Sigma_{V}^{\mu}\sim\mathcal{O}(\hbar^{0})$,
$\Sigma_{A}^{\mu}\sim\mathcal{O}(\hbar^{1})$.
We will only keep the terms up to the leading order of quantum corrections of
$\mathcal{O}(\hbar)$ in our following discussions. Given the formalism, our main task now is to derive the self-energies $\Sigma_{V/A}$ in
the 2 to 2 scattering process.

\section{Theoretical setup for an electron probe}\label{sec:general_setup}
Considering a hard-probe electron with 4-momentum $p$ emitted into
an electron plasma in equilibrium, we focus only on the $t$ channel
in the 2 to 2 scattering process,
\begin{equation}
	e^{-}(p)+e_{eq}^{-}(k)\leftrightarrow e^{-}(p^{\prime})+e_{eq}^{-}(k^{\prime}),   
\end{equation}
which leads to the leading-logarithmic contribution to the collision term. Since the probe electron and the scattered electron in the medium are treated as non-identical particles, we may neglect the
	$u$-channel scattering considered in Refs.~\cite{Hidaka:2016yjf,Hidaka:2017auj}. In principle, the Compton scattering should also yield a comparable contribution in the case for massless fermions. For simplicity, we assume the absence of thermalized photons in the plasma and ignore the related scattering processes. The positrons are also excluded.

\begin{figure}
	\begin{center}
		\includegraphics[width=0.45\hsize]{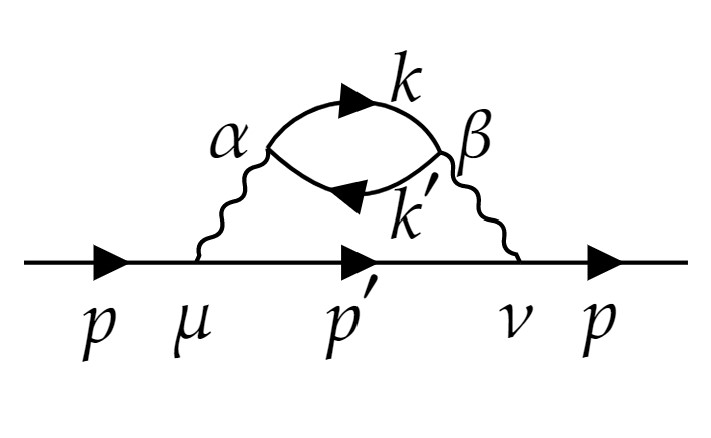}
	\end{center}
	\caption{The $t$ channel of the 2 to 2 scattering process, where the Roman indices denote the momenta of electrons and the greek indices represent the polarization of intermediate photons.}\label{fig:-channel-in}
\end{figure}

We now compute the self-energies shown in Fig.~\ref{fig:-channel-in},
\begin{eqnarray}
	-\Sigma^{\lessgtr}(p) & = & e^{2} \int_{p^{\prime},k^{\prime},k}\gamma_{\mu}S^{\lessgtr}(p^{\prime})\gamma_{\nu}\frac{-ig^{\mu\alpha}}{(p-p^{\prime})^{2}}\frac{-ig^{\nu\beta}}{(p-p^{\prime})^{2}}e^{2}\mathrm{Tr}\left[\gamma_{\alpha}S_{eq}^{\lessgtr}(k^{\prime})\gamma_{\beta}S_{eq}^{\gtrless}(k)\right]\nonumber \\
	&  & \times(2\pi)^{4}\delta^{(4)}(p+k-p^{\prime}-k^{\prime}),\label{eq:SE_Total}
\end{eqnarray}
where we have introduced 
\begin{equation}
	\int_{p}\equiv \int\frac{d^{4}p}{(2\pi)^{4}} .\;\;\;
\end{equation}
The Wigner functions in the trace in (\ref{eq:SE_Total}) are taken
as an in-equilibrium case because we mainly discuss the scattering between the
probe and thermalized electrons in a medium. We will discuss an alternative case in Sec.~ \ref{subsec:brief_summary}. Instead of using the resummed propagators of photons, we introduce the momentum cut-off for the free propagators to obtain leading-logarithmic results \cite{Arnold:2002zm,Li:2019qkf,Yang:2020hri}. Like the initial state, the final state of the hard probe  with momentum $p^{\prime}$ is not necessarily in equilibrium.

The fermionic self-energies can be written in terms of photonic self-energies
$\Pi_{\alpha\beta}^{\lessgtr}$,
\begin{eqnarray}
	\Sigma_{V,\mu}^{\lessgtr}(p) & = & e^{2}\int_{p^{\prime}}\left[\mathcal{V}^{\lessgtr,\rho}(p^{\prime})\widetilde{G}_{(\mu\rho)}^{\lessgtr}(q)-\mathcal{V}_{\mu}^{\lessgtr}(p^{\prime})\widetilde{G}_{\alpha}^{\lessgtr,\alpha}(q)\right],\\
	\Sigma_{A,\mu}^{\lessgtr}(p) & = & -e^{2}\int_{p^{\prime}}\left[\mathcal{A}^{\lessgtr,\rho}(p^{\prime})\widetilde{G}_{(\mu\rho)}^{\lessgtr}(q)-\mathcal{A}_{\mu}^{\lessgtr}(p^{\prime})\widetilde{G}_{\alpha}^{\lessgtr,\alpha}(q)+i\epsilon_{\mu\rho\alpha\beta}\mathcal{V}^{\lessgtr,\rho}(p^{\prime})\widetilde{G}^{\lessgtr,\alpha\beta}(q)\right].
	\label{eq:Fermion_SE_01} 
\end{eqnarray}
where we have defined 
\begin{equation}
	q^\mu \equiv p^\mu-p^{\prime \mu}.
\end{equation}
and introduced the photonic two-point Green function $\widetilde{G}_{\alpha\beta}^{\lessgtr}$
in one loop,
\begin{eqnarray}
	\frac{1}{2}\widetilde{G}^{\lessgtr,(\mu\nu)} & = & G^{\mu\alpha}(q)\frac{\Pi_{(\alpha\beta)}^{\lessgtr}(q)}{2}G^{\beta\nu,\dagger}(q),\\
	\frac{1}{2}\widetilde{G}^{\lessgtr,[\mu\nu]} & = & G^{\mu\alpha}(q)\frac{\Pi_{[\alpha\beta]}^{\lessgtr}(q)}{2}G^{\beta\nu,\dagger}(q).
\end{eqnarray}
Here, $G^{\mu\nu}(q)$ denotes the free photon propagator in the Feynman gauge, 
\begin{equation}
	G^{\mu\nu}(q)=\frac{-ig^{\mu\nu}}{q^{2}}.
\end{equation}
The symmetric and anti-symmetric components of photonic self-energies $\Pi_{\alpha\beta}^{\lessgtr}$ can be expressed in terms of Wigner functions as
\begin{eqnarray}
	\frac{1}{2}\Pi_{(\alpha\beta)}^{\lessgtr}(q) & = & 4e^{2}\int_{k^{\prime},k}\left[\mathcal{V}_{(\alpha,eq}^{\lessgtr}(k^{\prime})\mathcal{V}_{\beta,eq)}^{\gtrless}(k)-g_{\alpha\beta}\mathcal{V}_{eq}^{\lessgtr}(k^{\prime})\cdot\mathcal{V}_{eq}^{\gtrless}(k)\right]
	(2\pi)^{4}\delta^{(4)}(p+k-p^{\prime}-k^{\prime}),\label{eq:Photonic_SE_Sym}\\
	\frac{1}{2}\Pi_{[\alpha\beta]}^{\lessgtr}(q) & = & 4ie^{2}\int_{k,k^{\prime}}\epsilon_{\alpha\beta\delta\rho}\left[\mathcal{V}_{eq}^{\lessgtr,\rho}(k^{\prime})\mathcal{A}_{eq}^{\gtrless,\delta}(k)+\mathcal{A}_{eq}^{\lessgtr,\rho}(k^{\prime})\mathcal{V}_{eq}^{\gtrless,\delta}(k)\right](2\pi)^{4}\delta^{(4)}(p+k-p^{\prime}-k^{\prime}), \label{eq:Photonic_SE_Asym}
\end{eqnarray}
and we accordingly acquire the vector and axial self-energies for electrons,
\begin{eqnarray}
	\Sigma_{V}^{\lessgtr,\mu}(p) & = & 8e^{4}\int_{p^{\prime},k^{\prime},k}\frac{1}{(p-p^{\prime})^{4}}\left[\mathcal{V}_{eq}^{\gtrless}(k)\cdot\mathcal{V}^{\lessgtr}(p^{\prime})\mathcal{V}_{eq}^{\lessgtr,\mu}(k^{\prime})+\mathcal{V}_{eq}^{\gtrless,\mu}(k)\mathcal{V}_{eq}^{\lessgtr}(k^{\prime})\cdot\mathcal{V}^{\lessgtr}(p^{\prime})\right],\nonumber \\
	&  & \times(2\pi)^{4}\delta^{(4)}(p+k-p^{\prime}-k^{\prime})\\
	\Sigma_{A}^{\lessgtr,\mu}(p) & = & 8e^{4}\int_{p^{\prime},k^{\prime},k}\frac{1}{(p-p^{\prime})^{4}}(2\pi)^{4}\delta^{(4)}(p+k-p^{\prime}-k^{\prime})\nonumber \\
	&  & \times\left[ \mathcal{V}^{\lessgtr}(p^{\prime})\cdot\mathcal{A}_{eq}^{\gtrless}(k)\mathcal{V}_{eq}^{\lessgtr,\mu}(k^{\prime})+\mathcal{A}_{eq}^{\lessgtr,\mu}(k^{\prime})\mathcal{V}_{eq}^{\gtrless}(k)\cdot\mathcal{V}^{\lessgtr}(p^{\prime})\right.\nonumber \\
	&  & \;\;\;-\mathcal{A}_{eq}^{\gtrless,\mu}(k)\mathcal{V}_{eq}^{\lessgtr}(k^{\prime})\cdot\mathcal{V}^{\lessgtr}(p^{\prime})-\mathcal{A}_{eq}^{\lessgtr}(k^{\prime})\cdot\mathcal{V}^{\lessgtr}(p^{\prime})\mathcal{V}_{eq}^{\gtrless,\mu}(k)\nonumber \\
	&  & \;\;\;\left.-\mathcal{V}_{eq}^{\gtrless}(k)\cdot\mathcal{A}^{\lessgtr}(p^{\prime})\mathcal{V}_{eq}^{\lessgtr,\mu}(k^{\prime})-\mathcal{V}_{eq}^{\gtrless,\mu}(k)\mathcal{V}_{eq}^{\lessgtr}(k^{\prime})\cdot\mathcal{A}^{\lessgtr}(p^{\prime})\right] ,
\end{eqnarray}
where we have only kept the results up to $\mathcal{O}(\hbar)$.


\section{Collision kernels in hard-thermal-loop approximation\label{sec:Collision-kernels-in}}

\subsection{General discussion on the collision kernel in the local and global equilibrium}\label{subsec:General-discussion}
To compute the self-energies, we need to express the $\mathcal{V}_{\alpha}^{\gtrless}$ and $\mathcal{A}_{\alpha}^{\gtrless}$ in terms of the distribution functions. To avoid the confusion, we will denote the distribution functions of the medium as $\mathcal{N}_{V/A}(x,p)$ to distinguish them from $f_{V/A}(x,p)$ as the distribution functions of the probe. Following the findings in Refs.~\cite{Chen:2015gta, Hidaka:2017auj},
the distribution function for chiral fermions in the local equilibrium is given by
\begin{equation}
\mathcal{N}_{R/L,leq}^{<}(x,p)=\left[\exp\left(\beta u\cdot p-\beta\mu_{R/L}\pm\frac{\hbar}{2}\Omega_{\mu\nu}S_{(u)}^{\mu\nu}\right)+1\right]^{-1},
\end{equation}
where $\beta(x)=T(x)^{-1}$ is the inverse of temperature, $u^{\mu}=u^{\mu}(x)$
is the fluid velocity, $\mu_{R/L}(x)$ are the chemical potentials
for right/left-handed fermions, 
\begin{equation}
    \Omega_{\mu\nu}=\frac{\partial_{\mu}(\beta u_{\nu})-\partial_{\nu}(\beta u_{\mu})}{2},
\end{equation}
and is called the thermal vorticity tensor. Here the frame vector $n^{\mu}$ is chosen to be the fluid velocity $u^{\mu}$. We also introduce the chemical potentials for the vector and axial charges, 
\begin{equation}
    \mu_{V}=\frac{\mu_{R}+\mu_{L}}{2}\simeq\mu_{R/L} \sim  \mathcal{O}(1), \; \mu_{A}=\frac{\mu_{R}-\mu_{L}}{2}\sim \mathcal{O}(\hbar).
\end{equation}
Therefore,
the vector and axial distribution functions in local equilibrium read
\begin{eqnarray}
\mathcal{N}_{V,leq}^{<}(x,p) & = & \left[\exp\left(\beta u\cdot p-\beta\mu_{V}\right)+1\right]^{-1},\\
\mathcal{N}_{A,leq}^{<}(x,p) & = & -\frac{\hbar}{2}\mathcal{N}_{V,leq}^{<}(x,p)\mathcal{N}_{V,leq}^{>}(x,p)\Omega_{\mu\nu}S_{(u)}^{\mu\nu},
\end{eqnarray}
where we have kept the results up to $\mathcal{O}(\hbar)$. Now, from Eqs.~(\ref{eq:S<_R}, \ref{eq:def_V_A}), the vector and axial-vector components of Wigner functions become
\begin{equation}
\mathcal{V}_{leq}^{\lessgtr,\mu}(p)=2\pi\delta(p^{2})p^{\mu}\mathcal{N}_{V,leq}^{\lessgtr}(x,p), \label{eq:V_large_small_01}
\end{equation}
and 
\begin{eqnarray}
\mathcal{A}_{leq}^{<,\mu}(x,p) & = & 2\pi\hbar\delta(p^{2})\mathcal{N}_{V,leq}^{<}(x,p)\mathcal{N}_{V,leq}^{>}(x,p)\left\{S_{(u)}^{\mu\nu}\left[\partial_{\nu}(\beta\mu_{V})
 -p^{\alpha}\xi_{\alpha\nu}\right]+\frac{1}{4}\epsilon^{\mu\rho\alpha\nu}p_{\rho}\Omega_{\alpha\nu}\right\} \label{eq:A_leq_01}
\end{eqnarray}
with $\mathcal{A}_{leq}^{>,\mu}(x,p)=-\mathcal{A}_{leq}^{<,\mu}(x,p)$, where we have introduced the thermal shear tensor
\begin{equation}
    \xi_{\alpha\nu}=\frac{\partial_{\alpha}(\beta u_{\nu})+\partial_{\nu}(\beta u_{\alpha})}{2}\;.
\end{equation}

Equipped with the above results, we are able to calculate Eqs.~(\ref{eq:Photonic_SE_Sym},\,\ref{eq:Photonic_SE_Asym}),
which take the form,
\begin{eqnarray}
\frac{1}{2}\Pi_{(\alpha\beta)}^{\lessgtr}(q) & = & 4e^{2}\int\frac{d^{4}k}{(2\pi)^{2}}\delta[(q+k)^{2}]\delta(k^{2})\mathcal{N}_{V,leq}^{\lessgtr}(q+k)\mathcal{N}_{V,leq}^{\gtrless}(k)\nonumber \\
 &  & \;\;\;\;\times(2k_{\alpha}k_{\beta}+q_{\alpha}k_{\beta}+q_{\beta}k_{\alpha}-g_{\alpha\beta}q\cdot k),\\
\frac{1}{2}\Pi_{[\alpha\beta]}^{\lessgtr}(q) & = & \pm4\hbar ie^{2}\int\frac{d^{4}k}{(2\pi)^{2}}\epsilon_{\alpha\beta\delta\rho}\delta[(q+k)^{2}]\delta(k^{2})\mathcal{N}_{V,leq}^{\lessgtr}(q+k)\mathcal{N}_{V,leq}^{\gtrless}(k)I_{\Pi}^{\delta\rho},
\end{eqnarray}
where 
\begin{eqnarray}
I_{\Pi}^{\delta\rho} & = & \partial_{\nu}(\beta\mu_{V})[(q^{\delta}+k^{\delta})\mathcal{N}_{V,leq}^{\lessgtr}(k)S_{(u)}^{\rho\nu}(k)+k^{\delta}\mathcal{N}_{V,leq}^{\gtrless}(k+q)S_{(u)}^{\rho\nu}(q+k)]\nonumber \\
 &  & -\xi_{\gamma\nu}[k^{\delta}(q^{\gamma}+k^{\gamma})\mathcal{N}_{V,leq}^{\gtrless}(k+q)S_{(u)}^{\rho\nu}(q+k)+k^{\gamma}(q^{\delta}+k^{\delta})\mathcal{N}_{V,leq}^{\lessgtr}(k)S_{(u)}^{\rho\nu}(k)]\nonumber \\
 &  & +\frac{1}{2}\widetilde{\Omega}^{\rho\xi}[(q^{\delta}k_{\xi}+k^{\delta}k_{\xi})\mathcal{N}_{V,leq}^{\lessgtr}(k)+(q_{\xi}k^{\delta}+k^{\delta}k_{\xi})\mathcal{N}_{V,leq}^{\gtrless}(k+q)],
\end{eqnarray}
and we have also introduced 
\begin{equation}
\widetilde{\Omega}_{\rho\gamma}=\frac{1}{2}\epsilon_{\rho\gamma\lambda\sigma}\Omega^{\lambda\sigma}.
\end{equation}

The fermionic self-energies are thus given by
\begin{eqnarray}
\Sigma_{V,\mu}^{\lessgtr}(p) & = & (2\pi)^{3}8e^{4}\int_{p^{\prime},k}\frac{1}{q^{4}}\delta[(q+k)^{2}]\delta(k^{2})\delta(p^{\prime,2})f_{V}^{\lessgtr}(p^{\prime})\mathcal{N}_{V,leq}^{\lessgtr}(q+k)\mathcal{N}_{V,leq}^{\gtrless}(k)\nonumber \\
 &  & \times(2k_{\mu}p^{\prime}\cdot k+q_{\mu}p^{\prime}\cdot k+k_{\mu}p^{\prime}\cdot q),\\
\Sigma_{A,\mu}^{\lessgtr}(p) & = & -(2\pi)^{3}4e^{4}\int_{p^{\prime},k}\frac{1}{q^{4}}\delta[(q+k)^{2}]\delta(k^{2})\delta(p^{\prime,2})\mathcal{N}_{V,leq}^{\lessgtr}(q+k)\mathcal{N}_{V,leq}^{\gtrless}(k)\nonumber \\
 &  & \times[2f_{A}^{\lessgtr}(p^{\prime})(2k_{\mu}p\cdot k+q_{\mu}p\cdot k+k_{\mu}p\cdot q)\nonumber \\
 &  & +2\hbar S^{(u),\rho\nu}(p^{\prime})\partial_{\nu}f_{V}^{\lessgtr}(p^{\prime})(2k_{\mu}k_{\rho}+q_{\mu}k_{\rho}+q_{\rho}k_{\mu})\nonumber \\
 &  & \pm2\hbar(g_{\mu\delta}p_{\rho}^{\prime}-g_{\mu\rho}p_{\delta}^{\prime})f_{V}^{\lessgtr}(p^{\prime})I_{\Pi}^{\delta\rho}],
\end{eqnarray}
which yield the collision kernel in Eq. (\ref{eq:Vector_KE}), 
\begin{eqnarray}
\mathcal{C}_{V}[f_{V}] & \equiv & \Sigma_{V,\mu}^{<}(p)\mathcal{V}^{>,\mu}(p)-\Sigma_{V,\mu}^{>}(p)\mathcal{V}^{<,\mu}(p)
+\mathcal{O}(\hbar^2)
\nonumber \\
 & = & 8e^{4}\delta(p^{2})\int\frac{d^{4}q}{(2\pi)^{2}}\int\frac{d^{4}k}{(2\pi)^{2}}\frac{1}{q^{4}}\delta[(p-q)^{2}]\delta[(q+k)^{2}]\delta(k^{2})\nonumber \\
 &  & \;\;\times[2(k\cdot p)^{2}-2(k\cdot p)q\cdot k+2(p\cdot q)k\cdot p-(p\cdot k)q^{2}-(p\cdot q)(q\cdot k)]\nonumber \\
 &  & \;\;\times[f_{V}^{>}(p)f_V^{<}(p-q)\mathcal{N}_{V,leq}^{<}(k+q)\mathcal{N}_{V,leq}^{>}(k)\nonumber \\
 &  & \;\;-f_{V}^{<}(p)f_V^{>}(p-q)\mathcal{N}_{V,leq}^{<}(-k-q)\mathcal{N}_{V,leq}^{>}(-k)] +\mathcal{O}(\hbar^2), \label{eq:C_V_gen}
\end{eqnarray}
and another one in Eq. (\ref{eq:Axial_KE}),
\begin{eqnarray}
\mathcal{C}_{A}[f_{V},f_{A}] & = & \Sigma_{A,\mu}^{<}(p)\mathcal{V}^{>,\mu}(p)-\Sigma_{A,\mu}^{>}(p)\mathcal{V}^{<,\mu}(p)-(\Sigma_{V,\mu}^{<}\mathcal{A}^{>,\mu}-\Sigma_{V,\mu}^{>}\mathcal{A}^{<,\mu})\nonumber \\
 & = & -8e^{4}\delta(p^{2})\int\frac{d^{4}q}{(2\pi)^{2}}\int\frac{d^{4}k}{(2\pi)^{2}}\frac{1}{q^{4}}\delta[(p-q)^{2}]\delta[(q+k)^{2}]\delta(k^{2})\nonumber \\
 &  & \times \left\{ \frac{}{}\mathscr{A}_{1}f_{A}^{<}(p-q)-\mathscr{A}_{2}f_{A}^{<}(p)+\hbar[\frac{}{}\mathscr{B}_{1}^{\alpha}\partial_{\alpha}f_{V}^{<}(p-q)-\mathscr{B}_{2}^{\alpha}\partial_{\alpha}f_{V}^{<}(p)]\right.\nonumber \\
 &  & \left.+\hbar\left[-\partial_{\nu}(\beta\mu_{V})\mathscr{C}^{\nu}+\xi_{\gamma\nu}\mathscr{D}^{\gamma\nu}-\frac{1}{2}\widetilde{\Omega}^{\rho\xi}\mathscr{E}_{\rho\xi}\right]\right\} +\mathcal{O}(\hbar^2), \label{eq:C_A_gen}
\end{eqnarray}
where the explicit expressions of $\mathscr{A}_{i},\mathscr{B}_{i}^{\alpha},\mathscr{C}^{\nu},\mathscr{D}^{\gamma\nu},\mathscr{E}_{\rho\xi}$
are presented in App.~\ref{sec:app_express}.

As a sanity check for our results, we will show the detailed balance such that the collision kernels vanish when $f_{V/A}=f_{V/A,leq}=\mathcal{N}^<_{V/A,leq}$.
For convenience, one may change the integration variables $ k \rightarrow -k$ in $\mathcal{C}_V$. It is easy to check 
\begin{eqnarray}
\mathcal{C}_V[f_{V,leq}]=\Sigma_{V,\mu,leq}^{<}(p)\mathcal{V}_{leq}^{>,\mu}(p)-\Sigma_{V,\mu,leq}^{>}(p)\mathcal{V}_{leq}^{<,\mu}(p) & = & 0.\label{eq:geq_Vector_CK}
\end{eqnarray}
For the axial collision kernel, it is found
\begin{eqnarray}
\mathcal{C}_{A}[f_{V,leq};f_{A,leq}] & = & -(2\pi)^{4}8\hbar e^{4}\delta(p^{2})\int_{q,k}\delta[(p-q)^{2}]\delta[(q+k)^{2}]\delta(k^{2})\frac{1}{q^{4}}\nonumber \\
 &  & \times f_{V,leq}^{>}(k)f_{V,leq}^{>}(p)f_{V,leq}^{<}(p-q)f_{V,leq}^{<}(k+q)\nonumber \\
 &  & \times\{\partial_{\alpha}(\beta\mu_{V})I_{T}^{\alpha}+\xi_{\gamma\alpha}I_{shear}^{\gamma\alpha}\}+\mathcal{O}(\hbar^{2}), \label{eq:Axial_CK_leq}
\end{eqnarray}
where
\begin{eqnarray}
I_{T}^{\alpha} & = & S_{(u)}^{\mu\alpha}(p-q)I_{C1,\mu}-S_{(u)}^{\mu\alpha}(k)I_{C2,\mu}-S_{(u)}^{\mu\alpha}(q+k)I_{C3,\mu}-S_{(u)}^{\mu\alpha}(p)I_{C4,\mu}\nonumber \\
I_{shear}^{\gamma\alpha} & = & -S_{(u)}^{\mu\alpha}(p-q)(p^{\gamma}-q^{\gamma})I_{C1,\mu}+S_{(u)}^{\mu\alpha}(k)k^{\gamma}I_{C2,\mu}\nonumber \\
 &  & +S_{(u)}^{\mu\alpha}(q+k)(q^{\gamma}+k^{\gamma})I_{C3,\mu}+S_{(u)}^{\mu\alpha}(p)p^{\gamma}I_{C4,\mu},
\end{eqnarray}
and
\begin{eqnarray}
I_{C1,\mu} & = & 2p\cdot kk_{\mu}+p\cdot qk_{\mu}+p\cdot kq_{\mu},\nonumber \\
I_{C2,\mu} & = & q_{\mu}(p\cdot q+p\cdot k)-p_{\mu}(q^{2}+q\cdot k),\nonumber \\
I_{C3,\mu} & = & p\cdot kq_{\mu}-p_{\mu}q\cdot k,\nonumber \\
I_{C4,\mu} & = & (2p\cdot k-2q\cdot k+p\cdot q-q^{2})k_{\mu}+(p\cdot k-q\cdot k)q_{\mu}.
\end{eqnarray}
In Ref.~\cite{Hidaka:2017auj}, the authors have used the symmetry argument to prove that the axial collision kernel vanishes. Alternatively, since $I_{T}^{\alpha}$, $I_{shear}^{\gamma\alpha}$, and $f_{V,leq}$ should not contain the gradient terms as the higher-order corrections in $\hbar$, 
we can prove that integrating over the momenta $q,k$ leads to
\begin{eqnarray}
S^{\mu\alpha}_{(u)}(\bar{p})I_{Ci,\mu}\rightarrow S^{\mu\alpha}_{(u)}(p)(c_iu_{\mu}+\bar{c}_ip_{\mu})=0,
\end{eqnarray}
where $i=1,2,3,4$, $\bar{p}$ is an arbitrary momentum, and $c_i,\bar{c}_i$ are just unimportant factors.
Consequently, it is found
\begin{equation}
    \mathcal{C}_{A}[f_{V,leq};f_{A,leq}] =0. 
\end{equation}

Before ending this subsection, we would like to comment on the condition in global equilibrium. As mentioned above, in local equilibrium, the fluid velocity $u^\mu$, temperature $T$, and chemical potentials $\mu_{V,A}$ are functions of the spacetime. Differently, the global equilibrium means all the thermodynamic variables are constant and fluid velocity satisfies the Killing condition. Since we have already proved the collision kernels vanish in local equilibrium, it is obvious that the collision kernel is zero in global equilibrium. Notably, as shown above and in Ref.~\cite{Hidaka:2017auj}, the vanishing collision kernel for the QKT of massless fermions in local equilibrium is reached after integrating over the momenta of scattered particles, while the integrand in global equilibrium vanishes even without the integration.


\subsection{HTL approximation and the spin-polarization rate \label{subsec:HTL-approximation}}

Following Ref. \cite{Yang:2020hri}, we further adopt the HTL approximation
to analyze the collision kernels. As a common strategy, we assume
\begin{equation}
eT\ll q^{\mu}\ll T,k^{\mu},k^{\prime,\mu},p^{\mu},
\end{equation}
and we are mainly interested in acquiring the leading-logarithmic result $\sim e^{4}\ln e^{-1}$. That is, we will conduct the $|\bm q|$ expansion up to the terms contributing to logarithmic divergence of the collision term in the following calculations. For simplicity,
we further assume that $\beta\mu_{V}$ is negligible but $\partial_{\mu}(\beta\mu_{V})$
is finite, i.e., we will ignore $\beta\mu_{V}$ in the distribution functions when
performing the integral while keeping
$\partial_{\mu}(\beta\mu_{V})$. After lengthy yet straightforward calculations, we eventually obtain the collision
kernel in the HTL approximation. We only summarize the main steps
here. More details can be found in App. \ref{sec:Self-energy-of}.
For simplicity, we assume that the fluid cell is in its own local rest frame, i.e., $u^\mu \simeq (1, \boldsymbol{u})$ with $|\boldsymbol{u}|\ll1$.

The symmetric photonic self-energies are given by Eq. (\ref{eq:Symmetric_Lessor_SE_HTLA}),
\begin{equation}
\Pi_{(\alpha\beta)}^{<}(q)=\Pi_{(\alpha\beta)}^{>}(-q)=\frac{e^{2}}{\pi|\bm{q}|}\left(a_{2}t_{1,\alpha\beta}+\frac{|\bm{q}|^{2}\hat{q}^{2}}{4\beta}\Theta_{\alpha\beta}(q)\right)+\mathcal{O}(|\bm{q}|^{2}),
\end{equation}
where the explicit forms of $a_2$ and $t_1^{\alpha\beta}$ are shown in Eq.~(\ref{eq:Symmetric_Lessor_SE_HTLA}).
The anti-symmetric part can be written into different components, 
\begin{eqnarray}
\Pi_{[\alpha\beta]}^{<}(q)=-\Pi_{[\alpha\beta]}^{>}(-q) =  \Pi_{[\alpha\beta]}^{<,(\xi)}(q)+\Pi_{[\alpha\beta]}^{<,(\beta\mu)}(q)+\Pi_{[\alpha\beta]}^{<,(\omega)}(q)+\Pi_{[\alpha\beta]}^{<,(D\beta)}(q)+\mathcal{O}(|\bm{q}|),\label{eq:Pi_anti_sym}
\end{eqnarray}
where the upper labels, $\xi,\beta\mu,\omega,D\beta$ denote
the contributions proportional to thermal shear tensor $\xi_{\mu\nu}$,
$\partial_{\mu}(\beta\mu_{V})$, the kinetic vorticity $\omega^{\mu}=\epsilon^{\mu\nu\alpha\beta}u_{\nu}\partial_{\alpha}u_{\beta}/2$,
and $D\beta\equiv(u\cdot\partial)\beta$. The expressions for each
term are given by Eqs. (\ref{eq:eq:PI_xi}, \ref{eq:PI_beta_=00005Cmu},
\ref{eq:Pi_omega}, \ref{eq:Pi_Dbeta}). Although we adopt the Feynman gauge through this paper, as an indirect check, the photonic two-point Green function in the Coulomb gauge $G_{\mathrm{coul}}^{\mu\nu}(k)=i[\Theta^{\mu\nu}(k)+\hat{k}^{2}u^{\mu}u^{\nu}]/k^{2}$
reads
\begin{eqnarray}
\frac{1}{2}\widetilde{G}^{<,(\mu\nu)}(q) & = & \frac{\pi e^{2}T^{3}}{12|\bm{q}|q^{4}}\left[\Theta^{\mu\nu}(q)\hat{q}^{2}+\frac{1}{2}\hat{q}^{4}u^{\mu}u^{\nu}\right]\left(1-\frac{\beta\hat{q}_{0}|\bm{q}|}{2}\right)\nonumber \\
 &  & +\frac{e^{2}T|\bm{q}|}{72\pi q^{4}}\left[\Theta^{\mu\nu}(q)\frac{\hat{q}^{2}}{2}+\hat{q}^{4}u^{\mu}u^{\nu}\right]\hat{q}_{0}^{2}(3+\pi^{2}) \nonumber \\
 &  & +\frac{e^{2}T|\bm{q}|}{8\pi q^{4}}\left[\Theta^{\mu\nu}(q)\frac{\hat{q}^{2}}{2}-\hat{q}^{4}u^{\mu}u^{\nu}\right]+\mathcal{O}(|\bm{q}|^{-2}),\label{eq:G_larger_less_anti_sym}
\end{eqnarray}
where the first line agrees with the bookkeeping result \cite{Bellac:2011kqa}
and the remaining terms are higher-order corrections.


From Eq. (\ref{eq:Fermion_SE_01}), we can compute the fermionic self-energies in the HTL approximation, 
\begin{eqnarray}
\delta(p^{2})\Sigma_{V,\mu}^{<}(p) & = & \delta(p^{2})\frac{e^{4}}{(2\pi)^{3}}\int_{m_{D}}^{T}d|\bm{q}|\int_{-1}^{1}dz^{\prime}\int_{-\infty}^{+\infty}dq_{0}\frac{1}{2|\bm{p}|}\frac{1}{|\bm{q}|^{3}(\hat{q}_{0}^{2}-1)^{2}}\nonumber \\
 &  & \times\delta\left(q_{0}-|\bm{q}|z^{\prime}+\frac{1-z^{\prime,2}}{2}\frac{|\bm{q}|^{2}}{|\bm{p}|}\right)\left(1+z^{\prime}\frac{|\bm{q}|}{|\bm{p}|}+\frac{3z^{\prime,2}-1}{2}\frac{|\bm{q}|^{2}}{|\bm{p}|^{2}}\right)\nonumber \\
 &  & \times\left[I_{1,\mu}^{\Sigma}f_{V}^{<}(|\bm{p}|)-I_{\partial,\mu\alpha}^{\Sigma}\partial_{p_{\perp}}^{\alpha}f_{V}^{<}(|\bm{p}|)+I_{\partial,\mu\alpha\beta}^{\Sigma}\partial_{p_{\perp}}^{\alpha}\partial_{p_{\perp}}^{\beta}f_{V}^{<}(|\bm{p}|)+\mathcal{O}(|\bm{q}|^{3})\right],
\end{eqnarray}
where the expressions of $I_{V1,\mu}^{\Sigma},I_{\partial,\mu\alpha}^{\Sigma},I_{\partial,\mu\alpha\beta}^{\Sigma}$
are shown in Eq.~(\ref{eq:I_sigma_temp_02}) and
\begin{equation}
z^{\prime}\equiv\cos\langle\bm{p},\bm{q}\rangle=-\hat{p}_{\perp,\mu}\hat{q}_{\perp}^{\mu}=-\hat{p}_{\mu}\hat{q}_{\perp}^{\mu}.
\end{equation}
Here we introduce the thermal mass $m_D\sim eT$ as an infrared cut-off for $|\bm q|$ in order to extract the leading-logarithmic result.
Then, we integrate over $q^0$ and $z^\prime$ and obtain the expression for $\delta(p^{2})\Sigma_{V,\mu}^{\lessgtr}$,
\begin{eqnarray}
\delta(p^{2})\Sigma_{V,\mu}^{\lessgtr}(p) & = & -\frac{\pi^{2}e^{4}}{48\pi^{3}\beta^{2}}\delta(p^{2})\hat{p}_{\perp,\mu}\ln\frac{T}{m_{D}}\left\{ \left[\left(\frac{2}{\beta|\bm{p}|^{2}}+\frac{\pi^{2}-6}{6\pi^{2}}\beta\pm\frac{1}{|\bm{p}|}\right)\pm(\hat{p}_{\perp}\cdot\partial_{p_{\perp}})\right]f_{V}^{\lessgtr}(|\bm{p}|)\right.\nonumber \\
 &  & \left.+\left[-\frac{1}{\beta}(\partial_{p_{\perp}}\cdot\partial_{p_{\perp}})\pm\frac{\beta|\bm{p}|\pm2}{2\beta|\bm{p}|}\Theta_{\alpha\mu}(p)\partial_{p_{\perp}}^{\alpha}+\frac{1}{2\beta}\hat{p}_{\perp,(\alpha}\Theta_{\beta)\mu}(p)\partial_{p_{\perp}}^{\alpha}\partial_{p_{\perp}}^{\beta}\right]f_{V}^{\lessgtr}(|\bm{p}|)\right\} \nonumber \\
 &  & +\delta(p^{2})\Sigma_{\textrm{div},\mu}^{\lessgtr}(p),
\end{eqnarray}
where 
\begin{eqnarray}
\delta(p^{2})\Sigma_{\textrm{div},\mu}^{\lessgtr}(p) & = & -\frac{e^{4}}{8\pi\beta^{3}}\delta(p^{2})\hat{p}_{\perp,\mu}\int_{m_{D}}^{T}\frac{d|\bm{q}|}{|\bm{q}|^{3}}f_{V}^{\lessgtr}(|\bm{p}|)\nonumber \\
 &  & -\frac{\pi^{2}e^{4}}{48\pi^{3}\beta^{3}}\hat{p}_{\mu}\delta(p^{2})\int_{m_{D}}^{T}\frac{d|\bm{q}|}{|\bm{q}|^{3}}\int_{-1}^{1}\frac{dz^{\prime}}{(z^{\prime,2}-1)}\left\{ \left[2+|\bm{q}|^{2}a_{4}(z^{\prime},|\bm{p}|)\right]f_{V}^{\lessgtr}(|\bm{p}|)\right.\nonumber \\
 &  & +\left[\pm\frac{|\bm{q}|^{2}}{|\bm{p}|}(\beta|\bm{p}|\pm1)z^{\prime,2}(\hat{p}_{\perp}\cdot\partial_{p_{\perp}})+|\bm{q}|^{2}z^{\prime,2}(\hat{p}_{\perp}\cdot\partial_{p_{\perp}})^{2}\right.\nonumber \\
 &  & \left.\left.+\frac{|\bm{q}|^{2}}{2}\Theta_{\alpha\beta}(p)(z^{\prime,2}-1)\partial_{p_{\perp}}^{\alpha}\partial_{p_{\perp}}^{\beta}\right]f_{V}^{\lessgtr}(|\bm{p}|)+\mathcal{O}(|\bm{q}|^{3})\right\} .
\end{eqnarray}
and $a_{4}(z^{\prime},|\bm{p}|)=3z^{\prime,2}\frac{1}{|\bm{p}|^{2}}\pm\frac{1}{2|\bm{p}|}\beta+\frac{1}{6\pi^2}\beta^{2}z^{\prime,2}(3+\pi^{2})$. 
We note that the $\Sigma_{\textrm{div},\mu}^{\lessgtr}(p)$ are highly divergent in both collinear and infrared regimes, say, the terms $\sim \int_{-1}^{+1} dz^\prime (z^{\prime,2}-1)^{-1}$ corresponding to the former and $\sim \int_{m_D}^{T} d|\bm{q}||\bm{q}|^{-3} \sim m_D^{-2} \sim (eT)^{-2}$ corresponding to the latter. Here, we keep these divergent terms in the expression of $\Sigma_{V,\mu}^{\lessgtr}$. They will be exactly canceled in the collision kernel.

Recalling Eq.~(\ref{eq:Axial_KE}), we find that in the collision kernel $\mathcal{C}_{A}$,
the axial self-energies $\Sigma^{\lessgtr}_{A,\mu}$ are always combined with the vector component of Wigner functions as $\mathcal{V}^{\gtrless,\mu}\Sigma^{\lessgtr}_{A,\mu}$. To avoid the unnecessary
complexity, we compute the $\delta(p^{2})p\cdot\Sigma^{\lessgtr}_{A}$ instead
of the axial self-energy $\Sigma^{\lessgtr}_{A\mu}$. After a detailed calculation
shown in App. \ref{subsec:Anti-symmetric-Parts}, we eventually
obtain
\begin{eqnarray}
p^{\mu}\delta(p^{2})\Sigma_{A,\mu}^{\lessgtr}(p)[f_A,f_V] & = & \mp\frac{e^{4}}{16\pi^{3}}\delta(p^{2})|\bm{p}|\int_{m_{D}}^{T}d|\bm{q}|\frac{1}{|\bm{q}|^{3}}f_{A}^{<}(p)\frac{2\pi^{2}}{\beta^{3}}\nonumber \\
 &  & \mp\frac{e^{4}}{16\pi^{3}|\bm{p}|}\delta(p^{2})\ln\frac{T}{m_{D}}\left\{ |\bm{p}|^{2}f_{A}^{<}(p)\left[\frac{2\pi^{2}}{3\beta^{3}|\bm{p}|^{2}}\pm\frac{\pi^{2}}{3\beta^{2}|\bm{p}|}+\frac{\pi^{2}-6}{18\beta}\right]\right.\nonumber \\
 &  & -\frac{\pi^{2}}{3\beta^{3}}|\bm{p}|^{2}[(\partial_{p_{\perp}}\cdot\partial_{p_{\perp}})\mp\beta(\hat{p}_{\perp}\cdot\partial_{p_{\perp}})]f_{A}^{<}(p)+\hbar|\bm{p}|H_{3,\alpha}\partial_{p_{\perp}}^{\alpha}f_{V}^{\lessgtr}(p)\nonumber \\
 &  & \mp\hbar\frac{\pi^{2}}{12\beta^{2}}|\bm{p}|\epsilon^{\rho\alpha\nu\beta}\hat{p}_{\perp,\nu}u_{\beta}\partial_{p_{\perp},\rho}\partial_{\alpha}f_{V}^{<}(p)+\hbar\frac{\pi^{2}}{6\beta^{3}}\epsilon^{\rho\alpha\nu\beta}\hat{p}_{\perp,\rho}u_{\beta}\partial_{p_{\perp},\nu}\partial_{\alpha}f_{V}^{<}(p)\nonumber \\
 &  &\mp\hbar\frac{12+\pi^{2}}{36\beta^{2}}|\bm{p}|f_{V}^{\lessgtr}(p)\epsilon^{\kappa\nu\xi\lambda}u_{\kappa}\hat{p}_{\perp,\xi}\Omega_{\lambda\nu}\nonumber \\
 &  & \left.-\hbar\frac{\pi^{2}}{12\beta^{3}}|\bm{p}|\epsilon^{\rho\alpha\nu\beta}\hat{p}_{\perp,\nu}u_{\beta}\hat{p}_{\perp,(\gamma}g_{\lambda)\rho}\partial_{p_{\perp}}^{\lambda}\partial_{p_{\perp}}^{\gamma}\partial_{\alpha}f_{V}^{<}(p)
 \right\},\label{eq:Sigma_A_<_01}
\end{eqnarray}
where 
\begin{eqnarray}
H_{3,\alpha} & = & \frac{2\epsilon^{\kappa\xi\lambda\nu}}{\beta^3}\left[\frac{\pi^{2}}{72}g_{\alpha\xi}u_{\lambda}\xi_{\gamma\nu}(\hat{p}_{\perp}^{\gamma}\hat{p}_{\kappa}+3u^{\gamma}\hat{p}_{\perp,\kappa})+\frac{\beta\ln2}{4}u_{\lambda}g_{\alpha\kappa}\hat{p}_{\xi}\partial_{\nu}(\beta\mu_{V})\right.\nonumber \\
 &  & \left.+\frac{\pi^{2}}{48}\Omega_{\lambda\nu}\big(\Delta_{\alpha\kappa}\hat{p}_{\perp,\xi}+u_{\xi}(\hat{p}_{\perp,\alpha}\hat{p}_{\perp,\kappa}-g_{\alpha\kappa})\big)\right].
\end{eqnarray}
Finally, we get the vector collision kernel, 
\begin{eqnarray}
\mathcal{C}_{V}[f_{V}] & = & \frac{e^{4}\delta(p^{2})}{24\beta^{2}}\ln\frac{T}{m_{D}}\left[2f_{V}^{<}(p)f_{V}^{>}(p)+|\bm{p}|F(p)\hat{p}_{\perp,\alpha}\partial_{p_{\perp}}^{\alpha}f_{V}^{<}(p)-|\bm{p}|\frac{1}{\beta}(\partial_{p_{\perp}}\cdot\partial_{p_{\perp}})f_{V}^{<}(p)\right]+\mathcal{O}(\hbar^{2}),\nonumber \\
\label{eq:C_V_leq_01}
\end{eqnarray}
where 
\begin{equation}
F(p)\equiv f_{V}^{>}(p)-f_{V}^{<}(p)=1-2f_V(p).
\end{equation}
When combining $p\cdot\Sigma^{\lessgtr}_{A}(p)$ to compute the axial collision kernel $\mathcal{C}_{A}$, the collinear divergence and quadratic divergence from the soft-photon exchange in 
 $\Sigma_{A,\mu}^{<}(p)\mathcal{V}^{>,\mu}(p)-\Sigma_{A,\mu}^{>}(p)\mathcal{V}^{<,\mu}(p)$
in $\mathcal{C}_{A}$ and those in $\Sigma_{V,\mu}^{<}\mathcal{A}^{>,\mu}-\Sigma_{V,\mu}^{>}\mathcal{A}^{<,\mu}$ are exactly canceled.
At last, we obtain an axial collision kernel with only the logarithmic divergence regularized by the thermal mass as  
\begin{eqnarray}
\mathcal{C}_{A}[f_{V},f_{A}] & = & -\frac{e^{4}\delta(p^{2})}{8\pi^{2}|\bm{p}|}\ln\frac{T}{m_{D}}\left\{ \frac{2\pi^{2}}{3\beta^{2}}|\bm{p}|F(p)f_{A}^{<}(p)+\frac{\pi^{2}}{3\beta^{2}}|\bm{p}|^{2}F(p)[(\hat{p}_{\perp}\cdot\partial_{p_{\perp}})-\frac{1}{\beta}(\partial_{p_{\perp}}\cdot\partial_{p_{\perp}})]f_{A}^{<}(p)\right.\nonumber \\
 &  & -\frac{2\pi^{2}}{3\beta^{2}}|\bm{p}|^{2}f_{A}^{<}(p)(\hat{p}_{\perp}\cdot\partial_{p_{\perp}})f_{V}^{<}(p)+\hbar F(p)|\bm{p}|H_{3,\alpha}\partial_{p_{\perp}}^{\alpha}f_{V}^{<}(p)\nonumber \\
 &  & -\hbar\frac{\pi^{2}}{12\beta^{2}}F(p)|\bm{p}|\epsilon^{\rho\alpha\nu\beta}\hat{p}_{\perp,\nu}u_{\beta}\partial_{p_{\perp},\rho}\partial_{\alpha}f_{V}^{<}(p)+\hbar\frac{\pi^{2}}{6\beta^{3}}\epsilon^{\rho\alpha\nu\beta}\hat{p}_{\perp,\rho}u_{\beta}\partial_{p_{\perp},\nu}\partial_{\alpha}f_{V}^{<}(p)\nonumber \\
 &  & +\hbar\frac{\pi^{2}}{6\beta^{2}}\epsilon^{\mu\xi\lambda\kappa}p_{\lambda}u_{\kappa}\partial_{\xi}f_{V}^{<}(p)\partial_{p_{\perp},\mu}f_{V}^{<}(p)\nonumber \\
 &  & \left.-\hbar\frac{\pi^{2}}{12\beta^{3}}|\bm{p}|\epsilon^{\rho\alpha\nu\beta}\hat{p}_{\perp,\nu}u_{\beta}\hat{p}_{\perp,(\gamma}g_{\lambda)\rho}\hat{p}_{\perp,\lambda}\partial_{p_{\perp}}^{\lambda}\partial_{p_{\perp}}^{\gamma}\partial_{\alpha}f_{V}^{<}(p)\right\} +\mathcal{O}(\hbar^2).\label{eq:HTLA_CK_Axial_leq}
\end{eqnarray}
Note that some of the leading-logarithmic contributions from the lesser and greater parts of the self-energies also cancel each other and do not affect $\mathcal{C}_A$. 
Here many terms above in Eq.~(\ref{eq:HTLA_CK_Axial_leq}) could contribute to dynamical spin polarization. For simplicity, we may consider the scenario when there exists no initial axial charge and the spacetime gradient on $f_V$ is negligible, the spin-polarization rate here is then mostly governed by the $H_{3,\alpha}$ term in $\mathcal{C}_{A}$, which could be approximated as \footnote{We also neglect the spatial inhomogeneity of $f_A$.}
\begin{eqnarray}\label{eq:spin_pol_rate}
\Gamma_{A}(p)=\partial_0f_A(p)\approx \frac{\hbar e^{4}\ln e}{16\pi^{3}|\bm{p}|}F(p)H_{3,\alpha}\partial_{p_{\perp}}^{\alpha}f_{V}(p).
\end{eqnarray}
The polarization rate $\Gamma_A(p)$ is of importance to understand the angular-momentum transfer from spin-orbital interaction in the QKT. We will present more discussion on it in Sec. \ref{subsec:brief_summary}.

Now, we turn to the local-equilibrium limit. After inputting the equilibrium distribution
functions $f_{V,leq}$ into Eq.~(\ref{eq:C_V_leq_01}), one immediately finds
\begin{equation}
\mathcal{C}_{V}[f_{V,leq}]=0.
\end{equation}
On the other hand, by using the Schouten identity,
\begin{equation}
\epsilon^{\mu\alpha\rho\sigma}p^{\nu}+\epsilon^{\alpha\rho\sigma\nu}p^{\mu}+\epsilon^{\rho\sigma\nu\mu}p^{\alpha}+\epsilon^{\sigma\nu\mu\alpha}p^{\rho}+\epsilon^{\nu\mu\alpha\rho}p^{\sigma}=0,\label{eq:Schoten_id.}
\end{equation}
and with the help of 
$\partial_{p_{\perp},\alpha}\hat{p}_{\perp,\rho} \rightarrow \partial_{p_{\perp},i}\hat{p}_{\perp,j} =(-\delta_{ij}+\hat{p}_{\perp,i}\hat{p}_{\perp,j})/|\bm{p}|,\; (i,j)=1,2,3$,
we also obtain
\begin{equation}
\mathcal{C}_{A}[f_{V,leq},f_{A,leq}]=0,
\end{equation}
which is just up to our expectations. Although the collision kernels vanish due to the symmetry, in general,
as shown in Sec. \ref{subsec:General-discussion}, we emphasize that
it is highly non-trivial to check it with the QED interactions in
the HTL approximation. 

\subsection{A brief summary and discussions} \label{subsec:brief_summary}
In the previous section, we have computed the collision kernel in the HTL approximation and obtain the $C_V,C_A$ in a local-equilibrium medium. First, we summarize the QKT with the collision kernel (\ref{eq:Vector_KE}, \ref{eq:Axial_KE}) in HTL approximation. The vector and axial parts of QKT read\footnote{In principle, the axial kinetic equation incorporates an extra term proportional to $\hbar\partial_{\mu}(S^{\mu\nu}_{(u)}\mathcal{C}_{V,\nu}[f_V])$, where $\mathcal{C}_{V,\nu}[f_V]\equiv\Sigma^<_{V,\nu}f_V^>-\Sigma^>_{V,\nu}f_V^<$. However, since $\Sigma^{\lessgtr}_{V,\nu}\propto p_{\nu},u_{\nu}$ up to $\mathcal{O}(\hbar^0)$ in the effective power counting we adopt, it turns out that $\hbar S^{\mu\nu}_{(u)}\mathcal{C}_{V,\nu}[f_V]=0$.}
\begin{eqnarray}
	(p\cdot\partial) f_{V}^{<}(x,p) &=& \mathcal{C}_{V}^{\textrm{HTL}}[f_V] + \mathcal{O}(\hbar^2), \label{eq:CKT_vector_HTL_01} \\
	(p\cdot\partial) f_{A}^{<}(x,p) + \hbar\partial_{\mu}S^{\mu\nu}_{(u)}\partial_{\nu}f_{V}^{<}(x,p)& = &  \mathcal{C}_{A}^{\textrm{HTL}}[f_V,f_A] + \mathcal{O}(\hbar^2), 
	\label{eq:CKT_axial_HTL_01}
\end{eqnarray}
where $\mathcal{C}_{V, A}^{\textrm{HTL}}$ is related to the  $\mathcal{C}_{V,A}$ in Eqs.~(\ref{eq:C_V_leq_01}, \ref{eq:HTLA_CK_Axial_leq}) by
\begin{eqnarray}
	\mathcal{C}_{V,A}= 2\pi \delta(p^2) \mathcal{C}_{V, A}^{\textrm{HTL}}.
\end{eqnarray}
and the particles are on-shell. 

Given the collision kernels from the HTL approximation, we may estimate how fast the dynamical spin polarization of a probe compared to its thermalization when traversing a thermal medium in local equilibrium is. From Eq.~(\ref{eq:C_V_leq_01}), we may similarly estimate the interaction rate for $f_V$ as
\begin{eqnarray}
\Gamma_V(p)=\partial_0f_V(p)\approx \frac{e^{4}\ln e T^2}{|\bm p|}
\end{eqnarray} 
by further treating the rest terms in the bracket on the right-hand side of Eq.~(\ref{eq:C_V_leq_01}) as an $\mathcal{O}(1)$ quantity and omitting the overall numerical prefactor. On the other hand, approximating $F(p)|\bm p|\partial_{p_{\perp}}^{\alpha}f_{V}(p)/(8\pi^2)\sim \mathcal{O}(1)$ in Eq.~(\ref{eq:spin_pol_rate}) in the same fashion, we may estimate 
\begin{eqnarray}
\Gamma_{A}(p)\approx \frac{\hbar e^{4}\ln e}{|\bm{p}|^2}H_{3,\alpha},
\end{eqnarray}
and hence obtain the ratio
\begin{eqnarray}
\frac{\Gamma_{A}(p)}{\Gamma_{V}(p)}\approx \frac{\hbar H_{3,\alpha}}{T^2|\bm p|}\sim\mathcal{O}\left(\frac{\partial }{|\bm p|}\right),
\end{eqnarray}
where $\partial$ represents the gradient scale of the thermal medium. 
This result implies that the dynamical spin polarization for a probe could be much slower than its thermalization (for the vector distribution function) in certain cases, which is consistent with the finding from the NJL model \cite{Wang:2021qnt}. However, such a conclusion is based on the simplification in Eq.~(\ref{eq:spin_pol_rate}) and the omission of Compton scattering. In practice, large spacetime gradients on $f_V(p)$ that have been neglected could be present for an out-of-equilibrium probe even though the correction should be still within the valid regime for $\hbar$ expansion. The precise ratio will also depend on the initial condition for practical simulations of the full collision terms from Eqs.~(\ref{eq:C_V_leq_01}) and (\ref{eq:HTLA_CK_Axial_leq}).

For the future numerical simulations, we further simplify the QKT with collisions. One of the most important topics of spin polarization is to obtain the dynamical spin evolution equations near local equilibrium. It corresponds to taking the $f_V^<$ in Eqs.~(\ref{eq:CKT_vector_HTL_01}, \ref{eq:CKT_axial_HTL_01}) to be at local equilibrium. As shown in the previous section, $\mathcal{C}^{\textrm{HTL}}_V[f_{V,leq}]=0$. The evolution of $f_V ^<$ in Eq.~(\ref{eq:CKT_vector_HTL_01}) reduces to an ordinary Boltzmann equation near local equilibrium. On the other hand, Eq.~(\ref{eq:CKT_axial_HTL_01}) becomes 
\begin{eqnarray}
	(p\cdot\partial) f_{A}^{<}(x,p) + \hbar\partial_{\mu}S^{\mu\nu}_{(u)}\partial_{\nu}f_{V,leq}^{<}(x,p) =  \mathcal{C}_{A}^{\textrm{HTL}}[f_{V,leq},f_A] + \mathcal{O}(\hbar^2), \label{eq:spin_kinetic_eq} 
\end{eqnarray}
where
\begin{eqnarray}
	C_{A}^{\mathrm{HTL}}[f_{V,leq},f_{A}] & = & -\frac{e^{4}}{16\pi^{3}}\frac{\pi^{2}}{3\beta^{2}}\ln\frac{T}{m_{D}}\left\{ 2\left(f_{V,leq}^{>}(p)-f_{V,leq}^{<}(p)\right)+2|\bm{p}|\beta f_{V,leq}^{<}(p)f_{V,leq}^{>}(p)\right.\nonumber \\
	&  & \;\;\left.+|\bm{p}|\left[\left(f_{V,leq}^{>}(p)-f_{V,leq}^{<}(p)\right)\hat{p}_{\perp}\cdot\partial_{p_{\perp}}-\frac{1}{\beta}(\partial_{p_{\perp}}\cdot\partial_{p_{\perp}})\right]\right\} f_{A}^{<}(p)\nonumber \\
	&  & +\hbar\frac{e^{4}}{16\pi^{3}|\bm{p}|}\frac{\pi^{2}}{3\beta^{3}}\ln\frac{T}{m_{D}}S_{(u)}^{\alpha\nu}\Omega_{\alpha\nu}f_{V,leq}^{<}(p)f_{V,leq}^{>}(p)+\mathcal{O}(\hbar^{2}),\label{eq:Axial_Collision_Kernel_fV_leq}
\end{eqnarray}
which delineates the dynamical evolution for the spin. More precisely, the amplitude of the spin polarization is dynamically changed by $f_A$, while the direction is still fixed by the momentum for massless fermions here besides the non-dynamical part coming from the side-jump term in $\mathcal{A}^{<\mu}$.

In the previous section, we have assumed that both the vector and axial distribution functions for the medium are at local equilibrium. As an alternative scenario, we may consider the case when vector distribution functions for the medium is at local equilibrium, while the axial one $\mathcal{N}_A$ is not. We find there is an extra term, $\delta\Sigma_{A,\mu}^{\lessgtr}[\mathcal{N}_A,f_V]$, contributing to axial self-energies $p^\mu \delta(p^2)\Sigma_{A,\mu}^{\lessgtr}[\mathcal{N}_A,f_A,f_V]$, where we omit their $\mathcal{N}_{V,leq}$ dependence for brevity. It turns out that
\begin{eqnarray}
	p^\mu \delta(p^2)\Sigma_{A,\mu}^{\lessgtr}[\mathcal{N}_A,f_A,f_V] &=& p^\mu (p^2) \Sigma_{A,\mu}^{\lessgtr}[
	f_A, f_V ] + p^\mu \delta(p^2)\delta\Sigma_{A,\mu}^{\lessgtr}[\mathcal{N}_A^{<},f_V] , \label{eq:axial_sigma_02}\\
	p^{\mu}\delta(p^{2})\delta\Sigma_{A,\mu}^{\lessgtr}[\mathcal{N}_{A},f_V] & = & \frac{e^{4}}{8\pi^{3}}\delta(p^{2})f_{V}^{\lessgtr}(p) \int dq_{0}\int_{m_{D}}^{T}\frac{1}{|\bm{q}|}d|\bm{q}|\int_{-1}^{1}dz^{\prime}dz\int|\bm{k}|d|\bm{k}| dk_{0}\nonumber \\
	&  & \times \delta\left(q_{0}-|\bm{q}|z^{\prime}+\frac{1-z^{\prime,2}}{2}\frac{|\bm{q}|^{2}}{|\bm{p}|}\right)\delta\left(z-\frac{q^{2}+2q_{0}k_{0}}{2|\bm{q}||\bm{k}|}\right)\delta(k_{0}-|\bm{k}|)\nonumber \\
	&  & \times \left(1+\hat{p}_{\perp}\cdot\hat{k}_{\perp}\right)\frac{1}{z^{\prime,2}-1}\mathcal{N}_{A}^{<}(k)\left[\mathcal{N}_{V,leq}^{>}(k)-\mathcal{N}_{V,leq}^{<}(k)\right],
\end{eqnarray}
where $p^\mu \delta(p^2) \Sigma_{A,\mu}^{\lessgtr}[f_V,f_A ]$ is given by Eq.~(\ref{eq:Sigma_A_<_01}).  Inserting the above axial self-energies into the collision kernel $\mathcal{C}_{A}$, we find that the extra term $\delta\Sigma_{A,\mu}^{<}[\mathcal{N}_A,f_V]$ does not modify the $\mathcal{C}_{A}^{\textrm{HTL}}$ in Eq.~(\ref{eq:CKT_axial_HTL_01}). 
In this case, the axial kinetic equation (\ref{eq:spin_kinetic_eq}) holds. However, such a property may be subject to the HTL approximation.

\section{Near-equilibrium probe and relaxation-time approximation} \label{sec:relaxation_times}
In this section, we implement the relaxation-time approach 
to simplify the collision kernel for dynamical spin polarization of an electron probe approaching local equilibrium. Following the standard RTA by linearizing the collision term with respect to the fluctuation of $f_V$ and $f_A$ near local equilibrium, Eqs.~(\ref{eq:CKT_vector_HTL_01},\ref{eq:CKT_axial_HTL_01}) can be parameterized as, 
\begin{eqnarray}
(\hat{p}\cdot\partial)f_{V}^{<} & = & -\hat{\tau}^{-1}_{V,1}\delta f_{V},\label{eq:CKT_RTA_vector_01}\\
(\hat{p}\cdot\partial)f_{A}^{<}+\hbar|{\bm p}|^{-1}\partial_{\mu}S_{(u)}^{\mu\nu}\partial_{\nu}f_{V}^{<} & = & -\hat{\tau}_{A}^{-1}\delta f_{A}-\hat{\tau}^{-1}_{V,2}\delta f_{V}.\label{eq:CKT_RTA_axial_01}
\end{eqnarray}
Here, the $\hat{\tau}^{-1}_{V,1},\hat{\tau}^{-1}_A,\hat{\tau}^{-1}_{V,2}$ are the (inverse) relaxation-time operators and we introduce small deviations of the probe distribution functions from local equilibrium, 
\begin{eqnarray}
f_{V}(x,p) = f_{V,leq}(x,p)+\delta f_{V}(x,p), \;\;\;
f_{A}(x,p)  =  f_{A,leq}(x,p)+\delta f_{A}(x,p).
\end{eqnarray}
We also consider the gradient expansion and $\hbar$ expansion here. Then, Eq.~(\ref{eq:CKT_RTA_vector_01}) and Eq.~(\ref{eq:CKT_RTA_axial_01}) reduce to 
\begin{eqnarray}
\delta f_{V} &=& - \hat{\tau}_{V,1} (\hat{p}\cdot\partial)f_{V,leq}^{<}(x,p),
\\
\delta f_{A} &=&-\hat{\tau}_{A}(\hat{p}\cdot\partial)f_{A,leq}^{<}(x,p)-\hbar\hat{\tau}_{A}|\bm{p}|^{-1}\partial_{\mu}S_{(u)}^{\mu\nu}\partial_{\nu}f_{V,leq}^{<}(x,p)-\hat{\tau}_{A}\hat{\tau}_{V,2}^{-1}\delta f_{V},
\end{eqnarray}
up to $\mathcal{O}(\partial)$ for $\delta f_V$ and $\mathcal{O}(\partial^2)$ for $\delta f_A$, respectively.

From the collision kernels in Eqs.~(\ref{eq:CKT_vector_HTL_01},\,\ref{eq:CKT_axial_HTL_01}), we conduct the calculation and derive the explicit expression for the relaxation-time operators, 
\begin{eqnarray} \label{eq:tau_V_1_def}
 \hat{\tau}_{V,1}^{-1} & = & - \hat{\tau}_{A}^{-1} \nonumber \\
 &=& -\frac{e^{4}}{48\pi\beta^{2}{|\bm{p}|}}\ln\frac{T}{m_{D}}\left\{ 2[f_{V,leq}^{>}(p)-f_{V,leq}^{<}(p)]+2|\bm{p}|\beta f_{V,leq}^{<}(p)f_{V,leq}^{>}(p)\right. \nonumber \\
 &  & \left.+|\bm{p}|[f_{V,leq}^{>}(p)-f_{V,leq}^{<}(p)](\hat{p}_{\perp}\cdot\partial_{p_{\perp}})-|\bm{p}|\frac{1}{\beta}(\partial_{p_{\perp}}\cdot\partial_{p_{\perp}})\right\}, \\
 \hat{\tau}_{V,2}^{-1} & = & \frac{e^{4}\hbar}{48\pi\beta^{2}{|\bm{p}|}}\ln\frac{T}{m_{D}}\left[\Omega_{\mu\nu}S_{(u)}^{\mu\nu}\hat{a}_{5}+S_{(u)}^{\mu\nu}\hat{a}_{6}+\xi_{\gamma\nu}\hat{a}_{7}^{\gamma\nu}+\partial_{\nu}(\beta\mu_{V})\hat{a}_{8}^{\nu}+\Omega_{\lambda\nu}\hat{a}_{9}^{\lambda\nu}\right], \label{eq:tau_V_2_def}
\end{eqnarray}
where the operators $\hat{a}_5,\hat{a}_6,\hat{a}_7^{\gamma \nu}, \hat{a}_8^\nu, \hat{a}_9^{\lambda \nu}$ are shown in Eqs.~(\ref{eq:a_hat_5}, \ref{eq:a_hat_6}, \ref{eq:a_hat_7}, \ref{eq:a_hat_8}, \ref{eq:a_hat_9}). Note
$\hat{\tau}_{V,2}^{-1}\sim\mathcal{O}(\hbar)$ and $\hat{\tau}_{A}^{-1}\sim\hat{\tau}_{V,1}^{-1}\sim\mathcal{O}(1)$. 

In the standard RTA, one assume that these relaxation times are functions of $p$ instead of operators and immediately get the $\delta f_{V,A}$ \cite{Hidaka:2017auj,Hidaka:2018ekt}. By inserting the $\delta f_{V,A}$ into the modified Cooper-Frye formula \cite{Becattini:2013fla,Fang:2016uds}, we can get the additional contributions to the polarization pseudo-vector from interactions. However, as shown in Eqs.~(\ref{eq:tau_V_1_def},\ref{eq:tau_V_2_def}), the space and momentum derivatives are involved in the realistic relaxation times obtained from the field theory. Solving $\delta f_{V,A}$ analytically becomes rather challenging, which requires further studies in the future.


\section{Conclusions and outlook\label{sec:Conclusions-and-outlook}}
In this paper, we have investigated the collision kernels of QKT for a massless electron probing an local-equilibrium medium with QED-type interaction up to the leading-logarithmic order in the HTL approximation. The collision kernel for the axial kinetic equation delineating dynamical spin polarization is obtained in Eq.~(\ref{eq:HTLA_CK_Axial_leq}), from which we further extract the spin-polarization rate shown in Eq.~(\ref{eq:spin_pol_rate}). It turns out that the dynamical spin polarization of the probe is slower than its thermalization. Such an axial kinetic equation, denoted by Eq.~(\ref{eq:CKT_axial_HTL_01}), can be utilized for future simulations. Moreover, a simplified form for the electron probe approaching local equilibrium is further derived from the RTA, shown in Eq.~(\ref{eq:CKT_RTA_axial_01}), where the relaxation times in operator form are found. This kinetic equation will be useful for solving near-equilibrium corrections pertinent to interactions on the spin-polarization spectrum.

Our estimation of the spin-polarization rate for the massless fermions in a gauge theory complements that for massive fermions with contact interaction found in Ref.~\cite{Wang:2021qnt}. In general, the dynamical spin polarization is relatively slow, due to the suppression by the ratio of spacetime gradients to particle energy, compared to thermalization in the probe limit. It is hence desired to estimate the quantitative impact from non-equilibrium corrections upon the spin polarization. Even for a toy model, we have found the collision term could be rather complicated in the gauge theory. It infers great challenges to construct the practical collision term in QGP for the QKT, which is also implied by the complication of the spin-relaxation term in weakly coupled QGP \cite{Li:2019qkf,Yang:2020hri}. For phenomenological purpose, it could be enlightening to further employ our model to solve for near-local-equilibrium corrections, with suitable generalization to the massive case as Ref.~\cite{Yi:2021ryh}, and implement the hydrodynamic simulations to estimate the quantitative modifications on the local spin polarization. On the theoretical side, there also exists a puzzle that only the global-equilibrium solution of axial-vector Wigner functions is found from detailed balance of the QKT for massive fermions \cite{Weickgenannt:2020aaf,Wang:2020pej}, while the local-equilibrium corrections has been recently derived from other approaches in Refs.~\cite{Liu:2021uhn,Becattini:2021suc} for fermions with arbitrary mass. Some technical details in our study may also help with resolving this puzzle for the QKT with massive fermions, which is imperative to overcome in order to rigorously study non-equilibrium effects on spin polarization of massive fermions beyond the local-equilibrium contributions.

\begin{acknowledgments}
S.F. and S.P. are
supported by National Natural Science Foundation of China (NSFC) under
Grants No. 1207523 and No. 12135011. D.-L. Y. was supported by the
Ministry of Science and Technology, Taiwan under Grant No. MOST 110-2112-M-001-070-MY3.
\end{acknowledgments}

\appendix

\section{Expression for the coefficients in Eq.~(\ref{eq:C_A_gen})} \label{sec:app_express}

The coefficients 
$\mathscr{A}_{i},\mathscr{B}_{i}^{\alpha},\mathscr{C}^{\nu},\mathscr{D}^{\gamma\nu},\mathscr{E}_{\rho\xi}$ in Eq.~(\ref{eq:C_A_gen}) 
are 
\begin{eqnarray}
\mathscr{A}_{1} & = & \left[p\cdot k(2k\cdot p-2k\cdot q+2p\cdot q-q^{2})-(q\cdot k)p\cdot q\right]\nonumber \\
 &  & \times\left(f_{V}^{>}(p)\mathcal{N}_{V,leq}^{<}(k+q)\mathcal{N}_{V,leq}^{>}(k)+f_{V}^{<}(p)\mathcal{N}_{V,leq}^{>}(k+q)\mathcal{N}_{V,leq}^{<}(k)\right),\\
\mathscr{A}_{2} & = & \left[p\cdot k(2k\cdot p-2k\cdot q+2p\cdot q-q^{2})-(q\cdot k)p\cdot q\right]\nonumber \\
 &  & \times\left(f_{V}^{<}(p-q)\mathcal{N}_{V,leq}^{<}(k+q)\mathcal{N}_{V,leq}^{>}(k)+f_{V}^{>}(p-q)\mathcal{N}_{V,leq}^{>}(k+q)\mathcal{N}_{V,leq}^{<}(k)\right),\\
\mathscr{B}_{1}^{\alpha} & = & S^{(u),\rho\alpha}(p-q)\left(2p\cdot kk_{\rho}+p\cdot qk_{\rho}+p\cdot kq_{\rho}\right)\nonumber \\
 &  & \times\left[f_{V}^{>}(p)\mathcal{N}_{V,leq}^{<}(k+q)\mathcal{N}_{V,leq}^{>}(k)+\mathcal{N}_{V,leq}^{>}(k+q)\mathcal{N}_{V,leq}^{<}(k)f_{V}^{<}(p)\right],\\
\mathscr{B}_{2}^{\alpha} & = & S^{(u),\mu\alpha}(p)\left[(2p\cdot k-2q\cdot k+p\cdot q-q^{2})k_{\mu}+(p\cdot k-q\cdot k)q_{\mu}\right]\nonumber \\
 &  & \times\left[f_{V}^{>}(p-q)\mathcal{N}_{V,leq}^{>}(k+q)\mathcal{N}_{V,leq}^{<}(k)+f_{V}^{<}(p-q)\mathcal{N}_{V,leq}^{<}(k+q)\mathcal{N}_{V,leq}^{>}(k)\right],\\
\mathscr{C}^{\nu} & = & \left(q_{\rho}(p\cdot q+p\cdot k)-p_{\rho}(q^{2}+q\cdot k)\right)S_{(u)}^{\rho\nu}(k)\mathcal{N}_{V,leq}^{<}(k)\mathcal{N}_{V,leq}^{>}(k)\nonumber \\
 &  & \;\;\times\left[\mathcal{N}_{V,leq}^{<}(k+q)f_{V}^{>}(p)f_{V}^{<}(p-q)+\mathcal{N}_{V,leq}^{>}(k+q)f_{V}^{<}(p)f_{V}^{>}(p-q)\right]\nonumber \\
 &  & +(p\cdot kq_{\rho}-p_{\rho}q\cdot k)S_{(u)}^{\rho\nu}(q+k)\mathcal{N}_{V,leq}^{<}(k+q)\mathcal{N}_{V,leq}^{>}(q+k)\nonumber \\
 &  & \;\;\times\left[\mathcal{N}_{V,leq}^{<}(k)f_{V}^{<}(p)f_{V}^{>}(p-q)+\mathcal{N}_{V,leq}^{>}(k)f_{V}^{>}(p)f_{V}^{<}(p-q)\right],\\
\mathscr{D}^{\gamma\nu} & = & \left(q_{\rho}(p\cdot q+p\cdot k)-p_{\rho}(q^{2}+q\cdot k)\right)S_{(u)}^{\rho\nu}(k)k^{\gamma}\mathcal{N}_{V,leq}^{<}(k)\mathcal{N}_{V,leq}^{>}(k)\nonumber \\
 &  & \;\;\times\left[\mathcal{N}_{V,leq}^{<}(k+q)f_{V}^{>}(p)f_{V}^{<}(p-q)+\mathcal{N}_{V,leq}^{>}(k+q)f_{V}^{<}(p)f_{V}^{>}(p-q)\right]\nonumber \\
 &  & +(p\cdot kq_{\rho}-p_{\rho}q\cdot k)S_{(u)}^{\rho\nu}(q+k)(q^{\gamma}+k^{\gamma})\mathcal{N}_{V,leq}^{<}(k+q)\mathcal{N}_{V,leq}^{>}(q+k)\nonumber \\
 &  & \;\;\times\left[\mathcal{N}_{V,leq}^{<}(k)f_{V}^{<}(p)f_{V}^{>}(p-q)+\mathcal{N}_{V,leq}^{>}(k)f_{V}^{>}(p)f_{V}^{<}(p-q)\right],\\
\mathscr{E}_{\rho\xi} & = & \left(q_{\rho}k_{\xi}p\cdot k-p_{\rho}k_{\xi}q\cdot k+q_{\rho}k_{\xi}p\cdot q-p_{\rho}k_{\xi}q^{2}\right)\mathcal{N}_{V,leq}^{<}(k)\mathcal{N}_{V,leq}^{>}(k)\nonumber \\
 &  & \;\;\times\left[\mathcal{N}_{V,leq}^{<}(k+q)f_{V}^{>}(p)f_{V}^{<}(p-q)+\mathcal{N}_{V,leq}^{>}(k+q)f_{V}^{<}(p)f_{V}^{>}(p-q)\right]\nonumber \\
 &  & +\left(q_{\rho}k_{\xi}p\cdot k-p_{\rho}k_{\xi}q\cdot k-p_{\rho}q_{\xi}q\cdot k\right)\mathcal{N}_{V,leq}^{<}(k+q)\mathcal{N}_{V,leq}^{>}(q+k)\nonumber \\
 &  & \;\;\times\left[\mathcal{N}_{V,leq}^{<}(k)f_{V}^{<}(p)f_{V}^{>}(p-q)+\mathcal{N}_{V,leq}^{>}(k)f_{V}^{>}(p)f_{V}^{<}(p-q)\right].
\end{eqnarray}

\section{Expression for the operators in Eq.~(\ref{eq:tau_V_2_def})} \label{sec:expression_tau_V_2}

Here, we list the operators in Eq.~(\ref{eq:tau_V_2_def}),
\begin{eqnarray}
 \hat{a}_{5} & = & -\beta|\bm{p}|f_{V,leq}^{<}(p)f_{V,leq}^{>}(p)[f_{V,leq}^{>}(p)-f_{V,leq}^{<}(p)]+4f_{V,leq}^{<}(p)f_{V,leq}^{>}(p)\nonumber \\
 &  & +|\bm{p}|f_{V,leq}^{<}(p)f_{V,leq}^{>}(p)(\hat{p}_{\perp}\cdot\partial_{p_{\perp}}),\label{eq:a_hat_5} \\
\hat{a}_{6} & = & -\frac{1}{2}[f_{V,leq}^{>}(p)-f_{V,leq}^{<}(p)]\partial_{p_{\perp},\mu}\partial_{\nu}-\frac{1}{2\beta}\hat{p}_{\perp,(\gamma}g_{\lambda)\mu}\partial_{p_{\perp}}^{\lambda}\partial_{p_{\perp}}^{\gamma}\partial_{\nu}\nonumber \\
 &  & +\frac{1}{\beta{|\bm{p}|}}\partial_{p_{\perp},\nu}\partial_{\mu}+\left[\partial_{\nu}\left(\beta\mu_{V}\right)-p^{\gamma}\partial_{\nu}(\beta u_{\gamma})\right]f_{V,leq}^{<}(p)f_{V,leq}^{>}(p)\partial_{p_{\perp},\mu},\label{eq:a_hat_6} \\
\hat{a}_{7}^{\gamma\nu} & = & -\frac{1}{4\beta}[f_{V,leq}^{>}(p)-f_{V,leq}^{<}(p)]\epsilon^{\kappa\nu\xi\lambda}u_{\lambda}\left(-\frac{1}{3}\hat{p}_{\perp}^{\gamma}g_{\alpha\xi}\hat{p}_{\kappa}+u^{\gamma}\hat{p}_{\perp,\xi}g_{\alpha\kappa}\right)\partial_{p_{\perp}}^{\alpha},\label{eq:a_hat_7}\\
\hat{a}_{8}^{\nu} & = & \frac{3\ln2}{2\pi^{2}}[f_{V,leq}^{>}(p)-f_{V,leq}^{<}(p)]\epsilon^{\kappa\nu\xi\lambda}g_{\alpha\kappa}u_{\lambda}\hat{p}_{\xi}\partial_{p_{\perp}}^{\alpha},\label{eq:a_hat_8} \\
\hat{a}_{9}^{\lambda\nu} & = & \frac{1}{8\beta}[f_{V,leq}^{>}(p)-f_{V,leq}^{<}(p)]\epsilon^{\kappa\nu\xi\lambda}(g_{\alpha\kappa}\hat{p}_{\perp,\xi}-u_{\alpha}u_{\kappa}\hat{p}_{\perp,\xi}-g_{\alpha\kappa}u_{\xi}+u_{\xi}\hat{p}_{\perp,\alpha}\hat{p}_{\perp,\kappa})\partial_{p_{\perp}}^{\alpha}. \label{eq:a_hat_9} 
\end{eqnarray}

\section{Fermionic self energies in local equilibrium \label{sec:Self-energy-of}}

In this part we will compute the fermionic self-energies in the local-equilibrium
medium.

\subsection{Vector self-energies\label{subsec:Symmetric-Parts}}

Inserting the expression of $\mathcal{V}_{leq}^{\lessgtr,\mu}(p)$
from Eq.~(\ref{eq:V_large_small_01}) into Eq.~(\ref{eq:Photonic_SE_Sym}),
we get 
\begin{eqnarray}
\frac{1}{2}\Pi_{(\alpha\beta)}^{<}(q) & = & 4e^{2}\int\frac{d^{4}k}{(2\pi)^{2}}\delta\left[(q+k)^{2}\right]\delta(k^{2})\mathcal{N}_{V,leq}^{<}(k+q)\mathcal{N}_{V,leq}^{>}(k)\nonumber \\
 &  & \times\left(2k_{\alpha}k_{\beta}+q_{\alpha}k_{\beta}+q_{\beta}k_{\alpha}-g_{\alpha\beta}q\cdot k\right),\label{eq:Symmetric_Lessor_FSE_1}
\end{eqnarray}
where we define $q=p-p^{\prime}$ and the distribution function $f_{V}$
is at the equilibrium because the fermions are in the medium.


Before applying the HTL approximation, we first simplify the $\delta$
functions in Eq.~(\ref{eq:Delta_Function_1}). For simplicity,
we focus on the particles, whose energy is positive, and neglect the
negative-energy modes. We write the $\delta$ functions as, 
\begin{equation}
\delta\left[(q+k)^{2}\right]=\frac{1}{2|\bm{q}||\bm{k}|}\delta\left(z-\frac{q^{2}+2q_{0}k_{0}}{2|\bm{q}||\bm{k}|}\right),\qquad\delta(k^{2})=\frac{1}{2|\bm{k}|}\delta(k_{0}-|\bm{k}|),\label{eq:Delta_Function_1}
\end{equation}
where we have defined that 
\begin{equation}
z\equiv\cos\langle\bm{q},\bm{k}\rangle=\cos\langle\hat{q}_{\perp},\hat{k}_{\perp}\rangle=-\hat{q}_{\perp}\cdot\hat{k}_{\perp}.
\end{equation}

Also, we decompose the momentum as
\begin{equation}
k_{\alpha}=(u\cdot k)u_{\alpha}+\Delta_{\alpha\beta}k^{\beta}\simeq k_{0}u_{\alpha}+k_{\perp,\alpha},\;\;\;k_{\perp,\alpha}\equiv\Delta_{\alpha\beta}k^{\beta}.
\end{equation}
Here, we simplify the fluid velocity as $u^{\alpha}\simeq(1,\bm{u})$
and $|\bm{u}|\ll1$. We emphasize that the gradient of velocity
$\partial_{i}u_{j},\;(i,j=1,2,3)$ is finite and the chemical potential in the integral is neglected, while its gradient is kept, to get an analytic result.

In the following calculations, we also need to project the momentum
to the direction of $\hat{q}_{\perp}$.
Since we need to take the momentum integration at the end, e.g. in
Eq.~(\ref{eq:Symmetric_Lessor_FSE_1}), we can drop the last term
in the above decomposition due to the rotational symmetry with respect
to the polar axis, 
\begin{eqnarray}
k_{\alpha} & \to & k_{0}u_{\alpha}+z|\bm{k}|\hat{q}_{\perp,\alpha},
\nonumber \\
k_{\alpha}k_{\beta} & \to & k_{0}^{2}u_{\alpha}u_{\beta}+k_{0}|\bm{k}|z(u_{\alpha}\hat{q}_{\perp,\beta}+\hat{q}_{\perp,\alpha}u_{\beta})+z^{2}|\bm{k}|^{2}\hat{q}_{\perp,\alpha}\hat{q}_{\perp,\beta}+\frac{1}{2}|\bm{k}|^{2}\Theta_{\alpha\beta}(q)(z^{2}-1).\label{eq:Substitution_kk2}
\end{eqnarray}

Inserting Eq.~(\ref{eq:Delta_Function_1}) into Eq.~(\ref{eq:Symmetric_Lessor_FSE_1}) and applying the decomposition
(\ref{eq:Substitution_kk2})
yield 
\begin{eqnarray}
\frac{1}{2}\Pi_{(\alpha\beta)}^{<}(q) & = & \frac{e^{2}}{(2\pi)^{2}|\bm{q}|}\int d|\bm{k}|d\phi \mathcal{N}_{V,leq}^{<}(k)\mathcal{N}_{V,leq}^{>}(k) \nonumber \\
& & \times \left\{ \frac{}{}1-\beta q_{0}\mathcal{N}_{V,leq}^{>}(k)\right.\left.-\frac{1}{2}\beta^{2}q_{0}^{2}\mathcal{N}_{V,leq}^{>}(k)[\mathcal{N}_{V,leq}^{<}(k)-\mathcal{N}_{V,leq}^{>}(k)]\right\} \nonumber \\
 &  & \times\left\{ 2|\bm{k}|^{2}[u_{\alpha}u_{\beta}+a_{1}u_{(\alpha}\hat{q}_{\perp,\beta)}+a_{1}^{2}\hat{q}_{\perp,\alpha}\hat{q}_{\perp,\beta}+\frac{1}{2}\Theta_{\alpha\beta}(q)(a_{1}^{2}-1)] \right. \nonumber \\ 
 & & -|\bm{k}|g_{\alpha\beta}(u_{\rho}q_{0}+q_{\perp,\rho})(u^{\rho}+a_{1}\hat{q}_{\perp}^{\rho})\nonumber \\
 &  & \left.+|\bm{k}|q_{(\alpha}(u_{\beta)}+a_{1}\hat{q}_{\perp,\beta)})+\mathcal{O}(|\bm{q}|^{3})\frac{}{}\right\} ,
\end{eqnarray}
where $a_{1}=(q^{2}+2q_{0}|\bm{k}|)/(2|\bm{q}||\bm{k})$
and we expand $\mathcal{N}_{V,eq}^{<}(k+q)$ in the limit $|\bm{q}|\ll|\bm{k}|$,
\begin{eqnarray}
\mathcal{N}_{V,leq}^{<}(k+q) & = & \mathcal{N}_{V,leq}^{<}(k)-\beta q_{0}\mathcal{N}_{V,leq}^{<}(k)\mathcal{N}_{V,leq}^{>}(k)-\frac{\beta^{2}q_{0}^{2}}{2}\mathcal{N}_{V,leq}^{>}(k)\mathcal{N}_{V,leq}^{<}(k)\left(\mathcal{N}_{V,leq}^{<}(k)-\mathcal{N}_{V,leq}^{>}(k)\right) \nonumber\\
&& +\mathcal{O}(|\bm{q}|^{3}).\label{eq:f<(k+q)_Expansion}
\end{eqnarray}
Here, since we are only interested in the result up to the leading-log logarithmic
order $~e^{4}\ln{e}$, the expansion of $\Pi_{\alpha\beta}^{<}(q)$
up to $\mathcal{O}(|\textbf{q}|^{2})$ is sufficient.

Integrating over the $|\bm{k}|$ and the angle $\phi$ from $0$ to $2\pi$, we obtain
\begin{eqnarray}
\frac{1}{2}\Pi_{(\alpha\beta)}^{<}(q) & = & \frac{e^{2}}{2\pi|\bm{q}|}\left(a_{2}t_{1,\alpha\beta}+\frac{|\bm{q}|^{2}\hat{q}^{2}}{4\beta}\Theta_{\alpha\beta}(q)\right)+\mathcal{O}(|\bm{q}|^{2}),
\end{eqnarray}
where 
\begin{eqnarray}\label{eq:Symmetric_Lessor_SE_HTLA}
a_{2} & = & \frac{12\pi^{2}-6\pi^{2}\beta\hat{q}_{0}|\bm{q}|+\beta^{2}\hat{q}_{0}^{2}|\bm{q}|^{2}(-6+\pi^{2})}{36\beta^{3}}+\frac{|\bm{q}|^{2}\hat{q}^{2}}{4\beta},\nonumber \\
t_{1,\alpha\beta} & = & \hat{q}_{0}u_{(\alpha}\hat{q}_{\perp,\beta)}+\frac{1}{2}(3\hat{q}_{0}^{2}-1)\hat{q}_{\perp,\alpha}\hat{q}_{\perp,\beta}+\frac{1}{2}(\hat{q}_{0}^{2}-1)g_{\alpha\beta}+\frac{1}{2}(3-\hat{q}_{0}^{2})u_{\alpha}u_{\beta}.
\end{eqnarray}

By using the relation 
\begin{equation}
\frac{1}{2}\Pi_{(\alpha\beta)}^{>}(q)=\frac{1}{2}\Pi_{(\alpha\beta)}^{<}(-q),\label{eq:Symmetric_SE_Relation}
\end{equation}
it is straightforward to get the connection between the one-loop photon
propagator $\widetilde{G}^{<,(\mu\nu)}(q)$ and $\widetilde{G}^{>,(\mu\nu)}(-q)$,
\begin{eqnarray}
\frac{1}{2}\widetilde{G}^{<,(\mu\nu)}(q)=G^{\mu\alpha}(q)\frac{\Pi_{(\alpha\beta)}^{<}(q)}{2}G^{\beta\nu,\dagger}(q)=\frac{1}{2}\widetilde{G}^{>,(\mu\nu)}(-q).
\end{eqnarray}

From Eq. (\ref{eq:Fermion_SE_01}), we can compute the self-energy
for fermions, 
\begin{eqnarray}
\delta(p^{2})\Sigma_{V,\mu}^{<}(p) & = & \frac{e^{4}}{(2\pi)^{4}}\delta(p^{2})\int_{m_{D}}^{T}|\bm{q}|^{2}d|\bm{q}|\int_{-1}^{1}dz^{\prime}\int_{0}^{2\pi}d\phi^{\prime}\int_{-\infty}^{+\infty}dq_{0}\frac{\delta[(p-q)^{2}]f_{V}^{<}(p-q)}{|\bm{q}|^{5}(\hat{q}_{0}^{2}-1)^{2}}\nonumber \\
 &  & \left[2a_{2}|\bm{p}|\left(\frac{2\hat{q}_{0}-3z^{\prime}\hat{q}_{0}^{2}+z^{\prime}}{2}\hat{q}_{\perp,\mu}+\hat{p}_{\mu}\frac{\hat{q}_{0}^{2}-1}{2}+\frac{3-\hat{q}_{0}^{2}-2z^{\prime}\hat{q}_{0}}{2}u_{\mu}\right)\right.\nonumber \\
 &  & \left.-|\bm{p}|\frac{|\bm{q}|^{2}\hat{q}^{2}}{2\beta}(u_{\mu}+z^{\prime}\hat{q}_{\perp,\mu})+\mathcal{O}(|\bm{q}|^{3})\right],\label{eq:HTLA_Lessor_VectorSE}
\end{eqnarray}
where we have used $\int d^{4}p^{\prime}f(p^{\prime})=\int d^{4}qf(p-q)$,
and
\[
\hat{q}^{\rho}t_{1,\rho\sigma}=0.
\]
 Here, we have introduced 
\begin{equation}
z^{\prime}\equiv\cos\langle\bm{p},\bm{q}\rangle=-\hat{p}_{\perp,\mu}\hat{q}_{\perp}^{\mu}=-\hat{p}_{\mu}\hat{q}_{\perp}^{\mu}.
\end{equation}
Again, we have limited the particle with momentum $p$ to the positive-energy particle,
say, $p_{0}=|\bm{p}|>0$. 

Similar to Eqs. (\ref{eq:Delta_Function_1}, \ref{eq:f<(k+q)_Expansion}
), we apply the on-shell condition and get 
\begin{eqnarray}
\delta[(p-q)^{2}] & \simeq & \frac{1}{2|\bm{p}|}\delta\left(q_{0}-|\bm{q}|z^{\prime}+\frac{1-z^{\prime,2}}{2}\frac{|\bm{q}|^{2}}{|\bm{p}|}\right)\left(1+z^{\prime}\frac{|\bm{q}|}{|\bm{p}|}+\frac{3z^{\prime,2}-1}{2}\frac{|\bm{q}|^{2}}{|\bm{p}|^{2}}\right)+\mathcal{O}(|\bm{q}|^{3}),
\end{eqnarray}
and expand $f_{V}^{<}(p-q)$ with $q$, 
\begin{eqnarray}
f_{V}^{<}(p-q) & = & f_{V}^{<}(|\bm{p}|)-|\bm{q}|\hat{q}_{\perp}^{\alpha}\partial_{p_{\perp},\alpha}f_{V}^{<}(|\bm{p}|)+\frac{|\bm{q}|^{2}\hat{q}_{\perp}^{\alpha}\hat{q}_{\perp}^{\beta}}{2}\partial_{p_{\perp},\alpha}\partial_{p_{\perp},\beta}f_{V}^{<}(|\bm{p}|)+\mathcal{O}(|\bm{q}|^{3}),\label{eq:f_V_Expansion}
\end{eqnarray}
where we assume that $|q_{0}|,|\bm{q}|\ll|\bm{p}|$ and keep
the expansion up to $\mathcal{O}(|\bm{q}|^{2})$. Again, we decompose
the $q^{\mu}$ and drop the terms which will vanish after integrating
over momentum,
\begin{eqnarray}
\hat{q}_{\perp,\alpha} & \to & z^{\prime}\hat{p}_{\perp,\alpha},\nonumber\\
\hat{q}_{\perp,\alpha}\hat{q}_{\perp,\beta} & \to & z^{\prime,2}\hat{p}_{\perp,\alpha}\hat{p}_{\perp,\beta}+\frac{\Theta_{\alpha\beta}(p)}{2}(z^{\prime,2}-1),\nonumber \\
\hat{q}_{\perp,\alpha}\hat{q}_{\perp,\beta}\hat{q}_{\perp,\gamma} & \to & z^{\prime,3}\hat{p}_{\perp,\alpha}\hat{p}_{\perp,\beta}\hat{p}_{\perp,\gamma}-\frac{z^{\prime}(1-z^{\prime,2})}{2}\times\left(\hat{p}_{\perp,\alpha}\Theta_{\beta\gamma}(p)+\hat{p}_{\perp,\beta}\Theta_{\gamma\alpha}(p)+\hat{p}_{\perp,\gamma}\Theta_{\alpha\beta}(p)\right).\nonumber \\\label{eq:Sub_q_perp_1}
\end{eqnarray}

Then, Eq. (\ref{eq:HTLA_Lessor_VectorSE}) becomes 
\begin{eqnarray}
\delta(p^{2})\Sigma_{V,\mu}^{<}(p) & = & \delta(p^{2})\frac{e^{4}}{(2\pi)^{3}}\int_{m_{D}}^{T}d|\bm{q}|\int_{-1}^{1}dz^{\prime}\int_{-\infty}^{+\infty}dq_{0}\frac{1}{2|\bm{p}|}\frac{1}{|\bm{q}|^{3}(\hat{q}_{0}^{2}-1)^{2}}\nonumber \\
 &  & \times\delta\left(q_{0}-|\bm{q}|z^{\prime}+\frac{1-z^{\prime,2}}{2}\frac{|\bm{q}|^{2}}{|\bm{p}|}\right)\left(1+z^{\prime}\frac{|\bm{q}|}{|\bm{p}|}+\frac{3z^{\prime,2}-1}{2}\frac{|\bm{q}|^{2}}{|\bm{p}|^{2}}\right)\nonumber \\
 &  & \times\left[I_{1,\mu}^{\Sigma}f_{V}^{<}(|\bm{p}|)-I_{\partial,\mu\alpha}^{\Sigma}\partial_{p_{\perp}}^{\alpha}f_{V}^{<}(|\bm{p}|)+I_{\partial,\mu\alpha\beta}^{\Sigma}\partial_{p_{\perp}}^{\alpha}\partial_{p_{\perp}}^{\beta}f_{V}^{<}(|\bm{p}|)+\mathcal{O}(|\bm{q}|^{3})\right],
\end{eqnarray}
where 
\begin{eqnarray}
I_{V1,\mu}^{\Sigma} & = & 2a_{3}|\bm{p}|\left(\frac{2\hat{q}_{0}z^{\prime}-3z^{\prime,2}\hat{q}_{0}^{2}+z^{\prime,2}+\hat{q}_{0}^{2}-1}{2}\hat{p}_{\perp,\mu}+(1-z^{\prime}\hat{q}_{0})u_{\mu}\right)\nonumber \\
 &  & -\frac{1}{2\beta}|\bm{p}||\bm{q}|^{2}(\hat{q}_{0}^{2}-1)(u_{\mu}+z^{\prime,2}\hat{p}_{\perp,\mu}),\nonumber \\
I_{\partial,\mu\alpha}^{\Sigma} & = & \frac{|\bm{p}||\bm{q}|}{3\beta^{3}}(2\pi^{2}-\pi^{2}\beta\hat{q}_{0}|\bm{q}|)\left\{ \frac{2\hat{q}_{0}-3z^{\prime}\hat{q}_{0}^{2}+z^{\prime}}{2}\left[z^{\prime,2}\hat{p}_{\perp,\alpha}\hat{p}_{\perp,\mu}+\frac{1}{2}(z^{\prime,2}-1)\Theta_{\alpha\mu}(p)\right]\right.\nonumber \\
 &  & \left.+\frac{1}{2}(\hat{q}_{0}^{2}-1)z^{\prime}\hat{p}_{\mu}\hat{p}_{\perp,\alpha}+\frac{1}{2}(3-\hat{q}_{0}^{2}-2z^{\prime}\hat{q}_{0})z^{\prime}u_{\mu}\hat{p}_{\perp,\alpha}\right\} ,\nonumber \\
I_{\partial,\mu\alpha\beta}^{\Sigma} & = & |\bm{p}||\bm{q}|^{2}\frac{\pi^{2}}{3\beta^{3}}\times\left\{ \frac{1}{2}(\hat{q}_{0}^{2}-1)\hat{p}_{\mu}\left[z^{\prime,2}\hat{p}_{\perp,\alpha}\hat{p}_{\perp,\beta}+\frac{\Theta_{\alpha\beta}(p)}{2}(z^{\prime,2}-1)\right]\right.\nonumber \\
 &  & +\frac{1}{2}(2\hat{q}_{0}-3z^{\prime}\hat{q}_{0}^{2}+z^{\prime})\left[z^{\prime,3}\hat{p}_{\perp,\alpha}\hat{p}_{\perp,\beta}\hat{p}_{\perp,\mu}-\frac{z^{\prime}(1-z^{\prime,2})}{2}\hat{p}_{\perp,(\alpha}\Theta_{\beta\mu)}(p)\right]\nonumber \\
 &  & \left.+\frac{1}{2}(3-\hat{q}_{0}^{2}-2z^{\prime}\hat{q}_{0})u_{\mu}\left[z^{\prime,2}\hat{p}_{\perp,\alpha}\hat{p}_{\perp,\beta}+\frac{\Theta_{\alpha\beta}(p)}{2}(z^{\prime,2}-1)\right]\right\} , \label{eq:I_sigma_temp_02}
\end{eqnarray} 
with
\[
a_{3}=\frac{12\pi^{2}-6\pi^{2}\beta\hat{q}_{0}|\bm{q}|+\beta^{2}\hat{q}_{0}^{2}|\bm{q}|^{2}(3+\pi^{2})}{36\beta^{3}}-\frac{|\bm{q}|^{2}}{4\beta}.
\]
and $\hat{p}_{\perp,(\alpha}\Theta_{\beta\mu)}(p)=\hat{p}_{\perp,\alpha}\Theta_{\beta\mu}(p)+\hat{p}_{\perp,\beta}\Theta_{\mu\alpha}(p)+\hat{p}_{\perp,\mu}\Theta_{\alpha\beta}(p)$.

\subsection{Axial self-energies \label{subsec:Anti-symmetric-Parts}}

Next, we compute the axial self-energy. By using the same method as
in App. \ref{subsec:Symmetric-Parts}, inserting Eqs.~(\ref{eq:V_large_small_01},
\ref{eq:A_leq_01}) into Eq. (\ref{eq:Photonic_SE_Asym}) yields 
\begin{eqnarray}
\frac{1}{2}\Pi_{[\alpha\beta]}^{<}(q) & = & -\frac{1}{2}\Pi_{[\alpha\beta]}^{>}(-q)\nonumber \\
 & = & 4\hbar ie^{2}\epsilon_{\alpha\beta\delta\rho}\int\frac{d^{4}k}{(2\pi)^{2}}\delta[(q+k)^{2}]\delta(k^{2})\mathcal{N}_{V,leq}^{<}(q+k)\mathcal{N}_{V,leq}^{>}(k)\nonumber \\
 &  & \times\left\{ (q^{\delta}+k^{\delta})\mathcal{N}_{V,leq}^{<}(k)S_{(n)}^{\rho\nu}(k)[\partial_{\nu}(\beta\mu_{V})-k^{\gamma}\xi_{\gamma\nu}]\right.\nonumber \\
 &  & \;\;+k^{\delta}\mathcal{N}_{V,leq}^{>}(q+k)S_{(n)}^{\rho\nu}(q+k)[\partial_{\nu}(\beta\mu_{V})-(q^{\gamma}+k^{\gamma})\xi_{\gamma\nu}]\nonumber \\
 &  & \;\;\left.+\frac{1}{4}\epsilon^{\rho\xi\lambda\nu}\Omega_{\lambda\nu}[(q^{\delta}k_{\xi}+k^{\delta}k_{\xi})\mathcal{N}_{V,leq}^{<}(k)+(k^{\delta}q_{\xi}+k^{\delta}k_{\xi})\mathcal{N}_{V,leq}^{>}(q+k)]\right\} .
\end{eqnarray}
In the HTL approximation, we consider mainly the leading-logarithmic results
and we only need to keep the photon anti-symmetric self-energies up to
$\mathcal{O}(|\bm{q}|^{0})$. 

With the help of 
\begin{eqnarray}
\epsilon^{\delta\gamma\lambda\sigma}\Omega_{\lambda\sigma} & = & 2\beta(u^{\gamma}\omega^{\delta}-u^{\delta}\omega^{\gamma})+\epsilon^{\delta\gamma\lambda\sigma}u_{\lambda}D\beta_{\sigma},\nonumber \\
\epsilon_{\rho\sigma\beta\gamma}\omega^{\beta\gamma} & = & 2(u_{\sigma}\omega_{\rho}-u_{\rho}\omega_{\sigma})-2\epsilon_{\rho\sigma\beta\gamma}u^{\gamma}(u\cdot\partial)u^{\beta},
\end{eqnarray}
we decompose $\frac{1}{2}\Pi_{[\alpha\beta]}^{<}(q)$ as 
\begin{eqnarray}
\frac{1}{2}\Pi_{[\alpha\beta]}^{<}(q) & = & \frac{1}{2}\Pi_{[\alpha\beta]}^{<,(\xi)}(q)+\frac{1}{2}\Pi_{[\alpha\beta]}^{<,(\beta\mu)}(q)+\frac{1}{2}\Pi_{[\alpha\beta]}^{<,(\omega)}(q)+\frac{1}{2}\Pi_{[\alpha\beta]}^{<,(D\beta)}(q)+\mathcal{O}(|\bm{q}|).
\end{eqnarray}
The first term is given by
\begin{eqnarray}
\frac{1}{2}\Pi_{[\alpha\beta]}^{<,(\xi)}(q) & = & \frac{\hbar ie^{2}}{2\pi|\bm{q}|}\epsilon_{\alpha\beta\delta\rho}\epsilon^{\rho\nu\xi\lambda}\frac{u_{\lambda}}{2}\xi_{\gamma\nu}\ensuremath{\left[ a_5 H_{1,\;\xi}^{\gamma\delta}+\frac{1}{4\beta^{2}}|\bm{q}|H_{2,\;\xi}^{\gamma\delta}\right] },\label{eq:eq:PI_xi} \nonumber \\
\end{eqnarray}
where
\begin{eqnarray}
a_5 & =& -\frac{\pi^{2}}{6\beta^{3}}+\frac{3+\pi^{2}-9\ln2}{18\beta^{2}}|\bm{q}|\hat{q}_{0}+\frac{1}{4\beta^{2}}\hat{q}_{0}|\bm{q}| ,\nonumber \\
H_{1,\;\xi}^{\gamma\delta} & = & \hat{q}_{0}u^{\gamma}u^{\delta}\hat{q}_{\perp,\xi}+\hat{q}_{0}^{2}\hat{q}_{\perp,\xi}u^{(\delta}\hat{q}_{\perp}^{\gamma)}+\frac{\hat{q}^{2}}{2}u^{(\delta}\Theta_{\xi}^{\gamma)}+\hat{q}_{0}^{3}\hat{q}_{\perp}^{\gamma}\hat{q}_{\perp}^{\delta}\hat{q}_{\perp,\xi}\nonumber \\
 &  & +\frac{\hat{q}^{2}\hat{q}_{0}}{2}(\hat{q}_{\perp}^{(\gamma}\Theta_{\xi}^{\delta)}+\hat{q}_{\perp,\xi}\Theta^{\gamma\delta}),\nonumber \\
H_{2,\;\xi}^{\gamma\delta} & = & \left(\hat{q}_{0}^{3}-\frac{\hat{q}_{0}^{2}+3}{2}\right)u^{\gamma}u^{\delta}\hat{q}_{\perp,\xi}+\frac{1}{2}(-\hat{q}_{0}+3\hat{q}_{0}^{3})\hat{q}_{\perp,\xi}u^{[\delta}\hat{q}_{\perp}^{\gamma]}-2\hat{q}_{0}\hat{q}_{\perp,\xi}u^{\delta}\hat{q}_{\perp}^{\gamma}\nonumber \\
 &  & +\frac{1}{2}(1-3\hat{q}_{0}^{2})\hat{q}_{\perp}^{\gamma}\hat{q}_{\perp}^{\delta}\hat{q}_{\perp,\xi}+\frac{\hat{q}^{2}}{2}(-\hat{q}_{\perp}^{\gamma}\delta_{\xi}^{\delta}-\hat{q}_{0}u^{\gamma}\delta_{\xi}^{\delta}-\hat{q}_{\perp,\xi}g^{\gamma\delta}).
\end{eqnarray}
The term related to the gradient of $\beta\mu_{V}$ reads
\begin{eqnarray}
\frac{1}{2}\Pi_{[\alpha\beta]}^{<,(\beta\mu)}(q) & = & \frac{\hbar ie^{2}}{2\pi|\bm{q}|}\epsilon_{\alpha\beta\delta\rho}\epsilon^{\rho\nu\xi\lambda}\frac{u_{\lambda}}{2}\partial_{\nu}(\beta\mu_{V})\left\{ +\frac{1}{8\beta}|\bm{q}|\text{\ensuremath{\left[(2\hat{q}_{0}^{2}+1)\hat{q}^{\delta}\hat{q}_{\perp,\xi}+\hat{q}^{2}\hat{q}_{0}\hat{q}_{\perp,\xi}\hat{q}_{\perp}^{\delta}+\frac{3}{2}\hat{q}^{2}\hat{q}_{0}\Theta_{\xi}^{\delta}\right]}}\right.\nonumber \\
 &  & \;\;\left.+\left(\hat{q}_{0}u^{\delta}\hat{q}_{\perp,\xi}+\hat{q}_{0}^{2}\hat{q}_{\perp,\xi}\hat{q}_{\perp}^{\delta}+\frac{\hat{q}_{0}^{2}-1}{2}\Theta_{\xi}^{\delta}\right)\left(\frac{\ln2}{\beta^{2}}-\frac{2+\ln2}{3\beta}\hat{q}_{0}|\bm{q}|\pm\frac{1}{8\beta}|\bm{q}|\hat{q}_{0}\right)\right\} .\label{eq:PI_beta_=00005Cmu}
\end{eqnarray}
The last terms in $\frac{1}{2}\Pi_{[\alpha\beta]}^{<}(q)$ are
\begin{eqnarray}
\frac{1}{2}\Pi_{[\alpha\beta]}^{<,(\omega)}(q) & = & -\frac{\hbar ie^{2}}{4\pi|\bm{q}|}\epsilon_{\alpha\beta\delta\rho}\beta\left[\left(\hat{q}_{0}\hat{q}_{\perp}^{\rho}\omega^{\delta}+\frac{\hat{q}_{0}^{2}+1}{2}u^{\rho}\omega^{\delta}-\frac{3\hat{q}_{0}^{2}-1}{2}u^{\delta}\hat{q}_{\perp}^{\rho}\omega\cdot\hat{q}_{\perp}\right)\right.\nonumber \\
 &  & \;\;\;\;\left.\times\left(\frac{\pi^{2}}{6\beta^{3}}-|\bm{q}|\hat{q}_{0}\frac{\pi^{2}+3}{18\beta^{2}}\right)+|\bm{q}|\hat{q}_{\perp}^{\rho}\omega^{\delta}\frac{\hat{q}_{0}^{2}-1}{4\beta^{2}}\right],\label{eq:Pi_omega}\\
\frac{1}{2}\Pi_{[\alpha\beta]}^{<,(D\beta)}(q) & = & -\frac{\hbar ie^{2}}{8\pi|\bm{q}|}\epsilon_{\alpha\beta\delta\rho}\epsilon^{\delta\gamma\lambda\sigma}u_{\lambda}D\beta_{\sigma}\left[\left(\hat{q}_{0}u^{\rho}\hat{q}_{\perp,\gamma}+\frac{\hat{q}_{0}^{2}-1}{2}\delta_{\gamma}^{\rho}+\frac{3\hat{q}_{0}^{2}-1}{2}\hat{q}_{\perp}^{\rho}\hat{q}_{\perp,\gamma}\right)\right.\nonumber \\
 &  & \;\;\;\;\left.\times\left(\frac{\pi^{2}}{6\beta^{3}}-|\bm{q}|\hat{q}_{0}\frac{\pi^{2}+3}{18\beta^{2}}\right)+|\bm{q}|u^{\rho}\hat{q}_{\perp,\gamma}\frac{1-\hat{q}_{0}^{2}}{4\beta^{2}}\right].\label{eq:Pi_Dbeta}
\end{eqnarray}
In fact, the two terms above originate from the thermal-vorticity contribution to the anti-symmetric photonic self-energy as
	\begin{eqnarray}
		&  & \frac{1}{2}\Pi_{[\alpha\beta]}^{\lessgtr,(\Omega)}(q)\nonumber \\
		& = & \frac{\hbar ie^{2}}{8\pi|\bm{q}|}\epsilon_{\alpha\beta\delta\rho}\epsilon^{\rho\nu\xi\lambda}\Omega_{\lambda\nu}\left[\frac{|\bm{q}|\hat{q}^{2}}{4\beta^{2}}\left(-u^{\delta}\hat{q}_{\perp,\xi}+\hat{q}_{\perp}^{\delta}u_{\xi}\right)\right.\nonumber \\
		&  & \left.+\left(\pm\frac{\pi^{2}}{6\beta^{3}}-\frac{3+\pi^{2}}{18\beta^{2}}q_{0}\right)\times\left(u^{\delta}u_{\xi}+\hat{q}_{0}u^{\delta}\hat{q}_{\perp,\xi}+\hat{q}_{0}\hat{q}_{\perp}^{\delta}u_{\xi}+\hat{q}_{0}^{2}\hat{q}_{\perp}^{\delta}\hat{q}_{\perp,\xi}+\frac{\hat{q}_{0}^{2}-1}{2}\Theta_{\xi}^{\delta}\right)\right].
	\end{eqnarray}
Notice that in our axial collision kernel, we encounter terms like $-f^<_V(p-q)\times\frac{1}{2}\Pi^<_{\alpha\beta}(q)$ in $\Sigma_A^{<}$ and $f^>_V(p-q)\times\frac{1}{2}\Pi^<_{\alpha\beta}(-q)$ in $\Sigma_A^{>}$. Expanding them with respect to $q$, we get the leading-order terms like $-\frac{\pi^2}{6\beta^3}(-q\cdot\partial_p)f^<_V(p)$ in $\Sigma_A^{<}$ and $\frac{\pi^2}{6\beta^3}(-q\cdot\partial_p)f^>_V(p)$ in $\Sigma_A^{>}$, which just have the different signs and contribute to $H_{3,\alpha}$ terms as shown in Eq.~(\ref{eq:Sigma_A_<_01}). On the contrary, the other terms related to $\Omega_{\lambda\nu}$ terms in Eq.~(\ref{eq:Sigma_A_<_01}) are in the form of $-f^<_V(p)$ in $\Sigma_A^{<}$ and $-f^>_V(p)$ in $\Sigma_A^{>}$, which yield the same sign. 

Then, the corresponding one-loop propagator is 
\begin{eqnarray}
 &  & \frac{1}{2}\widetilde{G}^{<,[\mu\nu]}(q)=G^{\mu\alpha}(q)\frac{\Pi_{[\alpha\beta]}^{<}(q)}{2}G^{\beta\nu,\dagger}(q)=\frac{1}{q^{4}}\frac{1}{2}\Pi^{<,[\mu\nu]}(q),
\end{eqnarray}
with the relation, 
\begin{equation}
\frac{1}{2}\widetilde{G}^{>,[\mu\nu]}(q)=-\frac{1}{2}\widetilde{G}^{<,[\mu\nu]}(-q).
\end{equation}
Inserting Eq. (\ref{eq:Pi_anti_sym}) into the above expression, we
get the $\widetilde{G}^{<,[\mu\nu]}(q)$ with the Coulomb gauge shown
in Eq. (\ref{eq:G_larger_less_anti_sym}). 

According to Eq.~(\ref{eq:Axial_KE}), we find that in the collision kernel $\mathcal{C}_{A}$,
the axial self-energy $\Sigma_{A}^{\mu}$ is always combined with vector
one as $\mathcal{V}_{\mu}\Sigma_{A}^{\mu}$. To avoid the unnecessary
complexity, we compute the $p\delta(p^{2})\cdot\Sigma_{A}$ instead
of the axial self-energy $\Sigma_{A}^{\mu}$. After a long calculation
similar to those in App. \ref{subsec:Symmetric-Parts}, we finally
obtain 
\begin{eqnarray}
p^{\mu}\delta(p^{2})\Sigma_{A,\mu}^{\lessgtr}(p) & = & \mp\frac{e^{4}}{16\pi^{3}}\delta(p^{2})|\bm{p}|\int_{m_{D}}^{T}d|\bm{q}|\frac{1}{|\bm{q}|^{3}}f_{A}^{<}(p)\frac{2\pi^{2}}{\beta^{3}}\nonumber \\
 &  & \mp\frac{e^{4}}{16\pi^{3}|\bm{p}|}\delta(p^{2})\ln\frac{T}{m_{D}}\bigg\{ |\bm{p}|^{2}f_{A}^{<}(p)\left[\frac{2\pi^{2}}{3\beta^{3}|\bm{p}|^{2}}\pm\frac{\pi^{2}}{3\beta^{2}|\bm{p}|}+\frac{\pi^{2}-6}{18\beta}\right].\nonumber \\
 &  & -\frac{\pi^{2}}{3\beta^{3}}|\bm{p}|^{2}[(\partial_{p_{\perp}}\cdot\partial_{p_{\perp}})\pm(\hat{p}_{\perp}\cdot\partial_{p_{\perp}})]f_{A}^{<}(p)+\hbar|\bm{p}|H_{3,\alpha}\partial_{p_{\perp}}^{\alpha}f_{V}^{\lessgtr}(p)\nonumber \\
 &  & \mp\hbar\frac{\pi^{2}}{12\beta^{2}}|\bm{p}|\epsilon^{\rho\alpha\nu\beta}\hat{p}_{\perp,\nu}u_{\beta}\partial_{p_{\perp},\rho}\partial_{\alpha}f_{V}^{<}(|\bm{p}|)+\hbar\frac{\pi^{2}}{6\beta^{3}}\epsilon^{\rho\alpha\nu\beta}\hat{p}_{\perp,\rho}u_{\beta}\partial_{p_{\perp},\nu}\partial_{\alpha}f_{V}^{<}(p)\nonumber \\
 &  & -\hbar\frac{\pi^{2}}{12\beta^{3}}|\bm{p}|\epsilon^{\rho\alpha\nu\beta}\hat{p}_{\perp,\nu}u_{\beta}\hat{p}_{\perp,(\gamma}g_{\lambda)\rho}\partial_{p_{\perp}}^{\lambda}\partial_{p_{\perp}}^{\gamma}\partial_{\alpha}f_{V}^{<}(p)\bigg\}, \label{eq:Sigma_A_HTL}
\end{eqnarray}
where 
\begin{eqnarray}
H_{3,\alpha} & = & 2\epsilon^{\kappa\nu\xi\lambda}\left[-\frac{\pi^{2}}{6\beta^{3}}\frac{u_{\lambda}}{2}\xi_{\gamma\nu}\left(-\frac{1}{6}\hat{p}_{\perp}^{\gamma}g_{\alpha\xi}\hat{p}_{\kappa}+\frac{1}{2}u^{\gamma}\hat{p}_{\perp,\xi}g_{\alpha\kappa}\right)+\frac{\ln2}{\beta^{2}}\frac{u_{\lambda}}{2}\frac{g_{\alpha\kappa}}{2}\hat{p}_{\xi}\partial_{\nu}(\beta\mu_{V})\right.\nonumber \\
 &  & \left.+\frac{1}{4}\Omega_{\lambda\nu}\frac{\pi^{2}}{6\beta^{3}}\frac{g_{\alpha\kappa}\hat{p}_{\perp,\xi}-u_{\alpha}u_{\kappa}\hat{p}_{\perp,\xi}-g_{\alpha\kappa}u_{\xi}+u_{\xi}\hat{p}_{\perp,\alpha}\hat{p}_{\perp,\kappa}}{2}\right].
\end{eqnarray}

\bibliographystyle{h-physrev}
\bibliography{Main_CKT_Collision_published.bbl}
\end{document}